\documentclass{article}
\usepackage[utf8]{inputenc}
\usepackage{amssymb}
\usepackage{amsmath}
\usepackage{graphicx}
\usepackage{geometry}
\begin{document}
\setlength{\oddsidemargin}{0.0\textwidth}
\setlength{\evensidemargin}{0.0\textwidth}

\title{Variational methods for solving high dimensional quantum systems  }

\author{Daming Li\thanks{lidaming@sjtu.edu.cn} \\ School of Mathematical Sciences, Shanghai Jiao Tong University, Shanghai, 200240, China}

\maketitle
\date{}

\begin{abstract}
Variational methods are highly valuable computational tools for solving high-dimensional quantum systems. In this paper, we explore the effectiveness of three variational methods: the density matrix renormalization group (DMRG), Boltzmann machine learning, and the variational quantum eigensolver (VQE). We apply these methods to two different quantum systems: the fermi-Hubbard model in condensed matter physics and the Schwinger model in high energy physics.
To facilitate the computations on quantum computers, we map each model to a spin 1/2 system using the Jordan-Wigner transformation. This transformation allows us to take advantage of the capabilities of quantum computing.
We calculate the ground state of both quantum systems and compare the results obtained using the three variational methods. By doing so, we aim to demonstrate the power and effectiveness of these variational approaches in tackling complex quantum systems.
\end{abstract}

\maketitle


\section{Introduction}

Many exciting phenomena in quantum many-body systems is due to the interplay of quantum fluctuations and correlations. Celebrated examples are the superfluid Helium \cite{Feynman_262}\cite{Feynman_102}, the fractional quantum Hall effect
\cite{Tsui_1559}\cite{Laughlin_1395}, the Haldane phase in quantum spin
chains \cite{Haldane_464}\cite{Haldane_1153}, quantum spin liquids \cite{Anderson_153}, and high-temperature superconductivity \cite{Bednorz_189}.
The aim of theoretical physics is to understand the emergent properties for such challenging quantum many-body systems. The main difficulty in investigating quantum many-body problems is due to the fact that the Hilbert space spanned by the possible microstates grows exponentially with the system size.
To unravel the physics of microscopic model systems and to study the robustness of
quantum phases of matter, large scale numerical simulations are essential. The exact diagonization method is only possible for small many body systems. For large systems, efficient quantum Monte Carlo (QMC) methods can be applied.  In a large class of quantum many-body systems (e.g, fermionic degrees of freedom, geometric frustration), however, these QMC sampling techniques cannot be used effectively due to sign problem \cite{Daming Li_2016}. In this case, the variational methods have been shown to be a powerful tool to efficiently simulate quantum many-body systems.

The first kind of variational method is the density matrix renormalization group
(DMRG) method \cite{White_2863} which was originally conceived as an algorithm to
 study ground state properties of one-dimensional systems. The success of the DMRG method is based on the area law of the quantum ground state, and thus can be represented efficiently using matrix product states (MPS)
 \cite{Fannes_443}\cite{Hastings_2007}\cite{Schollwck_326}. This algorithm has been generalized to study
 two-dimensional systems \cite{Liang9214}, Abelian and non-Abelian symmetries \cite{McCulloch852}\cite{Singh050301}\cite{Singh115125}\cite{Singh195114}\cite{Weichselbaum327}, single-site optimization with density matrix perturbation \cite{White180403}\cite{Hubig155115}, hybrid real-momentum space representation \cite{Motruk155139}\cite{Ehlers125125}, and the development of real-space parallelization \cite{Stoudenmire155137}, continuous matrix product state (cMPS) \cite{Verstraete104}\cite{Haegeman_085118}. The cMPS can also used to study the Lieb-Liniger model \cite{Rincon_115107}\cite{Ganahl_220402}, the Gaudin-Yang model \cite{Chung_012004}, periodic bc atomtronics \cite{Draxler_045145}, 1+1 relative bose theories \cite{Tilloy_096007} etc..

The machine learning has been extensively used in physical sciences \cite{Carleo_045002}.
The second kind of variational method is based on machine learning, where the neutral network variational ansatz
efficiently represents highly correlated quantum
states and whose parameters are easily optimized by means
of the variational Monte Carlo (VMC) method \cite{Carleo_602}. Together with restricted Boltzmann machine (RBM) \cite{Zen101}-\cite{Choo121}, other network structures such as a feed-forward  \cite{Choo121}\cite{Cai97}\cite{Luo122}\cite{Sharir124}\cite{Kessler2021},
recurrent \cite{Hibat2020}\cite{Roth2020}, and convolutional neural networks \cite{Schmitt125}-\cite{Liang2204},
has been adopted.

The third kind of variational method is called the variational quantum eigensolver (VQE), which is extensively used in quantum calculation. This is a hybrid algorithm, where the quantum state is prepared by a quantum algorithm, which is  implemented according to a quantum circuit with many quantum gates. Classical computer is used to optimize the parameters used in quantum circuit.  Compared with the matrix product state and machine learning ansatz,
the variational state in VQE is realized by quantum algorithm, which can be implemented in the quantum hardware devices, and thus  exponential speedup over classical methods becomes possible.
The previous quantum algorithms to find the
ground state of a given Hamiltonian were based on adiabatic state preparation and quantum phase estimation
subroutines \cite{Abrams5165}\cite{Aspuru1704}, both of which have circuit depth
requirements beyond those available in the NISQ era.  Here we listed both the original
VQE architecture  \cite{Kandala2019} and some more advanced methods:
orthogonality constrained VQE \cite{Higgott156}, subspace expansion method \cite{McClean2017},
subspace VQE \cite{Nakanishi2019}, multistate contracted VQE \cite{Parrish2019}, adiabatically assisted VQE \cite{Saez2018}\cite{Cerezo2020}, accelerated VQE \cite{Wang2019}\cite{Wang010346}\cite{Wang09350}.
To study the dynamics of many-body quantum system, the conventional quantum Hamiltonian simulation algorithm such as the Trotter-Suzuki product formula \cite{Wang394} were used.
However, the circuit depth of standard Trotterization methods can rapidly exceed the coherence time of noisy quantum computers. This has led to recent proposals for variational approaches to dynamical simulation, including
 iterative variational algorithms \cite{Li2017}\cite{Yuan191}, imaginary time evolution \cite{McArdle2019},
general first order derivative equations with non-Hermitian Hamiltonians \cite{Endo15},
 adaptive ansatz to reduce the circuit depth \cite{Yao2011}\cite{Zhang2011},
weighted subspace VQE \cite{Heya1904}\cite{Nakanishi2019},
variational fast forwarding \cite{Cirstoiu2020}\cite{Gibbs2021}\cite{Khatri140}\cite{Commeau2009}\cite{Nakanishi2019}.

In this paper, these three variational methods are discussed in detail and are used to solve the ground state energy of the Fermi-Hubbard model and Schwinger model. The arrangement of the paper is as follows. The details of the three variational methods are given in sections \ref{DMRG}, \ref{23.11.7.1}, and \ref{23.11.7.5}. In sections \ref{23.9.19.1} and \ref{23.11.5.0}, the Fermi-Hubbard model and Schwinger model have been mapped to spin 1/2 systems using the Jordan-Wigner representation. In section \ref{23.9.19.2}, the ground state energy of the Fermi-Hubbard model is calculated using these three variational methods. The discussion is given in the final section \ref{23.11.8.0}.

\section{Density Matrix Renormalization group}\label{DMRG}
A general discrete space of the many-body quantum systems with $N$ sites has the structure
\begin{eqnarray}\label{23_5_7_0}
\mathcal{H}=\operatorname{span}\Big\{\left|s_0\right\rangle \otimes \cdots \otimes\left|s_{N-1}\right\rangle | s_i \in \mathcal{L}_i, i \in\{0, \ldots, N-1\}\Big\}
\end{eqnarray}
where $\mathcal{L}_i$ is the set of discrete quantum numbers at site $i$.
 $\{|s_i\rangle\}_{s_i\in \mathcal{L}_i}$ is the standard orthogonal basis for the local Hilbert space $\mathcal{H}_i$ corresponding to the $i$th site, $i=0,\cdots,N-1$. The whole Hilbert space $\mathcal{H}$ is the tensor product
of individual local Hilbert space: $\mathcal{H} = \otimes_{i=0}^{N-1} \mathcal{H}_i$. The standard orthogonal basis
  $\left|s_0 \cdots s_{N-1}\right\rangle \equiv \left|s_0\right\rangle \otimes \cdots \otimes\left|s_{N-1}\right\rangle$ in $\mathcal{H}$ is also called the computational basis.
 For the qubit systems, $\mathcal{L}_i=\{0,1\}$. For spin-1/2,
$\mathcal{L}_i=\{-1,1\}$, which is the two eigenvalues of Pauli matrix $Z$. For general spin-$S$ with integral or half integral $S=1/2,1,3/2,2,\cdots$,
$\mathcal{L}_i=\{-2S,-2S+2,\cdots,2S-2,2S\}$ and  $\mathcal{H}_i$ has the dimension $2S+1$. Sometime, the dimension of Hilbert space is reduced under some kind of restriction.
For example, the spin 1/2 system with the restriction $\sum_{i=0}^{N-1}
\sigma_i^z=0$, the dimension 16 of $\mathcal{H}$ is reduced to 6 if $N=4$.

The state $|\psi\rangle \in \mathcal{H}$ can be represented by
\begin{eqnarray}\label{23_5_7_1}
|\psi\rangle  = \sum_{s} \psi(s) |s\rangle =
\sum_{s_0,\cdots,s_{N-1}} \psi(s_0,\cdots,s_{N-1}) |s_0\cdots s_{N-1}\rangle
\end{eqnarray}
where the sum $s$ over all quantum numbers is assumed.
Since the computational basis is standard orthogonal, the wave function $\psi(s) = \langle s| \psi\rangle$
is the inner product of $| \psi\rangle$ and $| s\rangle$.

The Hamiltonian $H$ of many-body quantum system is a Hermitian operator from $d^N$ dimensional Hilbert space $\mathcal{H}$ into itself. We assume that the dimension $d$ of $\mathcal{H}_i$ are same for $i=0,\cdots,N-1$. The calculation of the spectra of $H$ is difficult due to the exponentially large dimension over $N$. For small $N$, exact diagonalization can be adopted since $H$ can be represented as the Hermitian matrix in the computation basis $|s\rangle$.

A general operator $\hat O$ in the basis $|s\rangle$ can be written as
\begin{eqnarray}\label{22_2_5_17_0}
\hat O  = \sum_{s,s^\prime} O^{s_0\cdots s_{N-1}}_{s_0^\prime\cdots s_{N-1}^\prime}
|s_0\cdots s_{N-1}\rangle \langle |s_{N-1}^\prime\cdots s_{0}^\prime |
\end{eqnarray}
Since the basis $|s\rangle$ is orthogonal to each other,
\begin{eqnarray}\label{22_2_5_17_1}
\hat O |s_0^\prime\cdots s_{N-1}^\prime \rangle  = \sum_{s}
 O^{s_0\cdots s_{N-1}}_{s_0^\prime\cdots s_{N-1}^\prime}
|s_0\cdots s_{N-1}\rangle
\end{eqnarray}
i.e., $\hat O |s^\prime\rangle  = \sum_{s} O^{s}_{s^\prime}|s\rangle$.
Apply this operator $\hat O$ to a general state $|\psi\rangle$ in (\ref{23_5_7_1}), one has
\begin{eqnarray}\label{22_2_5_17_2}
\hat  O |\psi \rangle  =
\sum_{s,s^\prime} O^{s}_{s^\prime}\psi(s^\prime) |s\rangle
\end{eqnarray}
The inner product between $\hat O |\psi \rangle $ and $  |\psi \rangle $ is
\begin{eqnarray}\label{22_2_5_17_3}
\langle \psi | \hat O |\psi \rangle  =
\sum_{s,s^\prime} O^{s}_{s^\prime}\psi(s^\prime) \overline{\psi(s)}
\end{eqnarray}
where $\overline{\psi(s)}$ denotes the complex conjugate of $\psi(s)$.

The wave function $\psi(s_0,\cdots,s_{N-1})$ can be regarded as the tensor
$\psi_{s_0\cdots s_{N-1}}$ of order $N$,  and thus the manipulation of the wave function
is reduced to be that of tensor $\psi_{s_0\cdots s_{N-1}}$. Similarly, $O^{s}_{s^\prime} = O^{s_0\cdots s_{N-1}}_{s_0^\prime\cdots s_{N-1}^\prime} $ is the tensor of order $2N$. The computational cost of
$O^{s}_{s^\prime}\psi(s^\prime) = O^{s_0\cdots s_{N-1}}_{s_0^\prime\cdots s_{N-1}^\prime}\psi_{s_0^\prime,\cdots,s_{N-1}^\prime}$ is $O(d^{2N})$ which is not acceptable. The trick of reducing computational cost is the decomposition of tensor and then truncation. A very important operation for tensor is the decomposition
as follows:
\begin{eqnarray}\label{22_2_5_17}
\psi_{s_0\cdots s_{N-1}}  =
 M^{[0]s_0} M^{[1]s_1} \cdots M^{[N-1]s_{N-1}}
\end{eqnarray}
where $M^{[i]s_i} = (M^{[i]s_i})_{\alpha_i\alpha_{i+1}} $ is the $\chi_{i} \times \chi_{i+1}$ matrix for fixed (physical) indices $s_i$, $i=0,\cdots,N-1$, i.e.,
 $M^{[i]}$ is a tensor of order 3. The index $\{\alpha_i\}_{i=0}^N$ are contracted in the multiplication
 of all $M^{[i]s_i}$, and $\alpha_0=\chi_0=\alpha_{N}=\chi_{N}=1$ since the contracted result $\psi_{s_0\cdots s_{N-1}}$ is a complex number.
Since the index $\{\alpha_i\}_{i=0}^N$ is not physical, they are called virtual indices.

The decomposition (\ref{22_2_5_17}) can be realized by $N-1$ steps of singular value decomposition (SVD) of matrices, which is given in detail in Appendix \ref{Appendix_0}. Inserting
(\ref{22_2_5_17}) into (\ref{23_5_7_1}), we get the matrix product state
\begin{eqnarray}\label{23_7_4_0}
|\psi\rangle  =
\sum_{s_0,\cdots,s_{N-1}}
M^{[0]s_0} M^{[1]s_1} \cdots M^{[N-1]s_{N-1}} |s_0\cdots s_{N-1}\rangle
\end{eqnarray}
If  $s_i=1,\cdots,d$ for $i=0,\cdots,N-1$, then $\chi_1=d$ and $\chi_{i+1}\leq d\chi_i$, $i=1,\cdots,N-2$, and thus $\chi_i \approx d^i$. The (exact) decomposition of (\ref{22_2_5_17}) will cause exponential disaster. Fortunately,
if the the state of interest is not highly entangled, the truncation in SVD will not cause much error for the approximation of this state.

As shown in Appendix \ref{Appendix_0}, if the tensor
$\psi_{s_0,\cdots,s_{N-1}}$ is normalized
\begin{eqnarray}\label{23_7_4_3_1}
\sum_{s_0,\cdots,s_{N-1}}|\psi_{s_{0},\cdots,s_{N-1}}|^2=1
\end{eqnarray}
 the $N$ tensors $\{M^{[i]}\}_{i=0}^{N-1}$ of order 3 satisfies left canonical condition
\begin{eqnarray}\label{23_7_4_3_2}
\sum_{s_i} ( M^{[i]s_i})^\dagger M^{[i]s_i}= \mathbb{I}_{\chi_{i+1}}  , \quad i=0,\cdots,N-1
\end{eqnarray}
where $\mathbb{I}_{i+1}$ is the $\chi_{i+1}\times \chi_{i+1}$ identity matrix. For a general tenor
$\psi_{s_0\cdots s_{N-1}}$ of order $N$, we have the left canonical form
\begin{eqnarray}\label{22_2_5_17_1}
\psi_{s_0\cdots s_{N-1}} =
 A^{[0]s_0} A^{[1]s_1} \cdots A^{[N-1]s_{N-1}}\Lambda^{[N]}
\end{eqnarray}
with the left canonical condition for the $N$ tensors $\{A^{[i]}\}_{i=0}^{N-1}$
\begin{eqnarray}\label{23_7_4_5}
\sum_{s_i} ( A^{[i]s_i})^\dagger A^{[i]s_i}= \mathbb{I}_{\chi_{i+1}}  , \quad i=0,\cdots,N-1
\end{eqnarray}
Here we intentionally introduce a number $\Lambda^{[N]}$ which becomes 1 if
$\psi(s_0,\cdots,s_{N-1})$ satisfies (\ref{23_7_4_3_1}).
In fact, this left canonical form
can also be obtained from the general matrix product form (\ref{22_2_5_17}) where $M^{[i]}$ does not satisfies
(\ref{23_7_4_3_2}). This can be realized by $N-1$ steps of SVD. First,
\begin{eqnarray}\label{22_2_5_36}
 (M^{[0]s_0})_{\alpha_0,\alpha_1} =   \sum_{k} U_{(\alpha_0s_0), k} S_{k}(V^{\dagger})_{k ,\alpha_1}
\end{eqnarray}
where $U$ and $V$ are unitary matrices and $\{S_k\}$ are the singular values.
Then $(A^{[0]s_0})_{\alpha_0, k} = U_{(\alpha_0s_0), k}$ satisfies the left canonical condition. The remaining  part $S_{k}(V^{\dagger})_{k,\alpha_1}$ is combined with $ (M^{[1]s_1})_{\alpha_1,\alpha_2}$ to get
$\sum_{\alpha_1}S_{k}(V^{\dagger})_{k ,\alpha_1}(M^{[1]s_1})_{\alpha_1,\alpha_2}$ and use SVD for this matrix again, we can find $A^{[1]s_1}$ which satisfies the left canonical condition. Continuing this process, we can get the canonical
form (\ref{22_2_5_17_1}).

Similarly, if the SVD are implemented from the right hand side, the general tensor has the right canonical form
\begin{eqnarray}\label{23_7_4_6}
\psi_{s_0\cdots s_{N-1}} =
\Lambda^{[0]} B^{[0]s_0} B^{[1]s_1} \cdots B^{[N-1]s_{N-1}}
\end{eqnarray}
with the right canonical condition for the tensor $\{B^{[i]}\}_{i=0}^{N-1}$ of order 3
\begin{eqnarray}\label{23_7_4_7}
\sum_{s_i}  B^{[i]s_i}(B^{[i]s_i})^\dagger = \mathbb{I}_{\chi_{i}}  , \quad i=0,\cdots,N-1
\end{eqnarray}

If the SVD are implemented from both sides, they will meet at some bonds, e.g.,
\begin{eqnarray}\label{23_7_4_8}
\psi_{s_0\cdots s_{N-1}} =
\Lambda^{[0]} A^{[0]s_0}  A^{[1]s_1} \Lambda^{[2]} B^{[2]s_2} \cdots B^{[N-1]s_{N-1}} \Lambda^{[N]}
\end{eqnarray}
We call this the mixed canonical form. Here $\Lambda^{[2]}$ is a $\chi_2\times \chi_{2}$ matrix.

A full decomposition is
\begin{eqnarray}\label{23_7_4_10}
\psi_{s_0\cdots s_{N-1}} =
\Lambda^{[0]} \Gamma^{[0]s_0} \cdots \Lambda^{[N-1]} \Gamma^{[N-1]s_{N-1}} \Lambda^{[N]}
\end{eqnarray}
where $\Gamma^{[i]s_i}$ is the $\chi_i\times \chi_{i+1}$ matrix and
$\Lambda^{[i]}$ is the $\chi_i\times \chi_{i}$ diagonal matrix. Obviously, these three canonical forms
are related to each other
\begin{eqnarray}\label{23_7_4_11}
A^{[i]s_i} =  \Lambda^{[i]} \Gamma^{[i]s_i} = \Lambda^{[i]} B^{[i]s_i} (\Lambda^{[i+1]})^{-1}, \quad
B^{[i]s_i} =  \Gamma^{[i]s_{i}} \Lambda^{[i+1]} = (\Lambda^{[i]})^{-1} A^{[i]s_i}\Lambda^{[i+1]}
\end{eqnarray}
for $i=0,\cdots,N-1$.

Assume that $O^{s}_{s^\prime} = O^{s_0\cdots s_{N-1}}_{s_0^\prime\cdots s_{N-1}^\prime} $ has the decomposition
\begin{eqnarray}\label{22_2_7_30}
O =  v^L W^{[0]}\cdots W^{[N-1]} v^R
\end{eqnarray}
where $W^{[i]} = (W^{[i]}_{s_is_i^\prime})_{\alpha_i\alpha_{i+1}} $ is the tensor of order 4 with $1\leq s_i,s_i^\prime\leq d$, $1\leq\alpha_i\leq \chi_i$ and  $1\leq\alpha_{i+1}\leq \chi_{i+1}$, $v^L$ and $v^R$ are the
$1\times \chi_1$ and $\chi_N\times 1$ matrices, respectively. For example, the Hamiltonian
\begin{eqnarray}\label{22_2_7_40}
  \hat      H &=& - \sum_{i=0}^{N-2} J X_i X_{i+1}  - \sum_{i=0}^{N-1} g Z_i \nonumber \\
        & = & - J \sum_{i=0}^{N-2} \mathbb{I} \otimes \cdots  \otimes \mathbb{I}\otimes \underbrace{X}_{\text{site }i}  \otimes \underbrace{X}_{\text{site }i+1} \otimes \mathbb{I} \otimes\cdots  \otimes \mathbb{I}\nonumber \\
 && - g\sum_{i=0}^{N-1} \mathbb{I} \otimes \cdots  \otimes \mathbb{I}\otimes \underbrace{Z}_{\text{site }i}  \otimes  \mathbb{I} \otimes \cdots  \otimes \mathbb{I}
\end{eqnarray}
can be written in the form in (\ref{22_2_7_30}) with
$$ v^L = (1,0,0), \quad v^R = (0,0,1)^T, \quad  W^{[i]}=\left(\begin{array}{ccc}
1 & X & -gZ\\
0 & 0 &  -J X  \\
0 & 0 &  1  \\
\end{array}\right), \quad i=0,\cdots,N-1 $$
Here $X$, $Y$ and $Z$ are the Pauli $x$, $y$ and $z$ matrices, respectively.

From the decompositions in (\ref{22_2_5_17}) and (\ref{22_2_7_30}), one has
\begin{eqnarray}\label{23_7_5_3}
 O^{s_0\cdots s_{N-1}}_{s_0^\prime\cdots s_{N-1}^\prime}\psi_{s_0^\prime,\cdots,s_{N-1}^\prime}  &=&   v^L_{\alpha_0} (W^{[0]}_{s_0s_0^\prime})_{\alpha_0\alpha_1}\cdots (W^{[N-1]}_{s_{N-1}s_{N-1}^\prime})_{\alpha_{N-1}\alpha_{N}} v^R_{\alpha_N}  \nonumber \\
 &&  \hspace{0.5cm} (M^{[0]s_0^\prime})_{\beta_0\beta_1}  \cdots (M^{[N-1]s_{N-1}^\prime})_{\beta_{N-1}\beta_N}
\end{eqnarray}
The inner product $O |\psi \rangle $ with $ |\psi \rangle $ in (\ref{22_2_5_17_3}) is a scalar
\begin{eqnarray}\label{23_7_5_5}
\langle \psi |\hat O |\psi \rangle  &= &  O^{s_0\cdots s_{N-1}}_{s_0^\prime\cdots s_{N-1}^\prime}\psi_{s_0^\prime,\cdots,s_{N-1}^\prime}
   \overline{\psi_{s_0,\cdots,s_{N-1}}} \nonumber \\
    &=&   v^L_{\alpha_0} (W^{[0]}_{s_0s_0^\prime})_{\alpha_0\alpha_1}\cdots (W^{[N-1]}_{s_{N-1}s_{N-1}^\prime})_{\alpha_{N-1}\alpha_{N}} v^R_{\alpha_N}  \nonumber \\
 && \hspace{0.5cm} (M^{[0]s_0^\prime})_{\beta_0\beta_1}  \cdots (M^{[N-1]s_{N-1}^\prime})_{\beta_{N-1}\beta_N} \\
 && \hspace{0.5cm}
   \overline{(M^{[0]s_0})}_{\gamma_0\gamma_1} \cdots \overline{(M^{[N-1]s_{N-1}})}_{\gamma_{N-1}\gamma_N} \nonumber
\end{eqnarray}
The summation over one virtual index $\alpha_0$ and two physical indices $s_0$ and $s_0^\prime$ of $v^L_{\alpha_0}(W^{[0]}_{s_0s_0^\prime})_{\alpha_0\alpha_1}$,
$(M^{[0]s_0^\prime})_{\beta_0\beta_1}  $ and $\overline{(M^{[0]s_0})}_{\gamma_0\gamma_1}$ is a tensor of order 3 with
virtual indices $\alpha_1$, $\beta_1$ and $\gamma_1$. Note that $\beta_0=\gamma_0=1$.

If the operator $\hat O$ is local and $\psi_s$ is a canonical form, the expression in (\ref{23_7_5_5}) can be simplified.
For example, $\psi_s$ has the mixed canonical form (\ref{23_7_4_8}), $\hat O$ is the operator acting on local site 2 and site 3, i.e., $  (W^{[i]}_{s_is_i^\prime})_{\alpha_i,\alpha_{i+1}} =  \delta_{s_i,s_i^\prime}\delta_{\alpha_i,\alpha_{i+1}}$ for $i\neq 2,3$,
\begin{eqnarray}\label{23_7_5_7}
\langle \psi |\hat O |\psi \rangle
    &=&   v^L_{\alpha_0} (W^{[0]}_{s_0s_0^\prime})_{\alpha_0\alpha_1}(W^{[1]}_{s_1s_1^\prime})_{\alpha_1\alpha_2}
  \ \ \ \ \  (W^{[2]}_{s_2s_2^\prime})_{\alpha_2\alpha_3}(W^{[3]}_{s_3s_3^\prime})_{\alpha_3\alpha_4}   \cdots (W^{[N-1]}_{s_{N-1}s_{N-1}^\prime})_{\alpha_{N-1}\alpha_{N}} v^R_{\alpha_N}  \nonumber \\
 && \hspace{0.6cm} (A^{[0]s_0^\prime})_{\beta_0\beta_1}(A^{[1]s_1^\prime})_{\beta_1\beta_2}  \ \  (\Lambda^{[2]} B^{[2]s_2^\prime})_{\beta_2\beta_3}(B^{[3]s_{3}^\prime})_{\beta_{3}\beta_4}  \cdots (B^{[N-1]s_{N-1}^\prime})_{\beta_{N-1}\beta_N}
  \\
 && \hspace{0.6cm} (\overline{A^{[0]s_0}})_{\gamma_0\gamma_1}(\overline{A^{[1]s_1}})_{\gamma_1\gamma_2} \ \  (\Lambda^{[2]} \overline{ B^{[2]s_2}})_{\gamma_2\gamma_3}
(\overline{B^{[3]s_{3}}})_{\gamma_{3}\gamma_4} \ \cdots
 (\overline{B^{[N-1]s_{N-1}}})_{\gamma_{N-1}\gamma_N} \nonumber \\
    &=&   v^L_{\alpha_0} \delta_{\alpha_0,\alpha_1}\delta_{\alpha_1,\alpha_2}
    (W^{[2]}_{s_2s_2^\prime})_{\alpha_2\alpha_3}(W^{[3]}_{s_3s_3^\prime})_{\alpha_3\alpha_4}   \cdots \delta_{\alpha_{N-1},\alpha_{N}} v^R_{\alpha_N}  \nonumber \\
 && \hspace{1.0cm} \delta_{\beta_2,\gamma_2}   (\Lambda^{[2]} B^{[2]s_2^\prime})_{\beta_2\beta_3}(B^{[3]s_{3}^\prime})_{\beta_{3}\beta_4}  \delta_{\beta_{4},\gamma_{4}}
 \nonumber \\
 && \hspace{1.9cm}    (\Lambda^{[2]} \overline{ B^{[2]s_2}})_{\gamma_2\gamma_3}
(\overline{B^{[3]s_{3}}})_{\gamma_{3}\gamma_4}   \nonumber \\
    &=&   v^L_{\alpha_2}
    (W^{[2]}_{s_2s_2^\prime})_{\alpha_2\alpha_3}(W^{[3]}_{s_3s_3^\prime})_{\alpha_3\alpha_4}v^R_{\alpha_4}   \nonumber \\
 && \hspace{0.4cm}  (\Lambda^{[2]} B^{[2]s_2^\prime})_{\beta_2\beta_3}(B^{[3]s_{3}^\prime})_{\beta_{3}\beta_4}
 \nonumber  \\
 && \hspace{0.4cm}    (\Lambda^{[2]} \overline{ B^{[2]s_2}})_{\beta_2\gamma_3}
(\overline{B^{[3]s_{3}}})_{\gamma_{3}\beta_4}   \nonumber
\end{eqnarray}

The ground state of Hamiltonian $\hat H$ can be calculated by DMRG algorithm \cite{White_2863}, which is a variational method by minimizing $\langle \psi | \hat H  |\psi \rangle$ over the tensor $\psi_s$ of order $N$. Here we always require that $\sum_s|\psi_s|^2 =1$.  Since $\psi_s$ has the structure of matrix product state, this global minimization can be realized by local minimization step by step. For example, $\hat H$ and $\psi_s$ have a form (\ref{22_2_7_30}), and (\ref{23_7_4_8}), respectively,  then $\langle \psi | \hat H |\psi \rangle$ has a form (\ref{23_7_5_7}) can be written as
\begin{eqnarray}\label{23_7_5_8}
\langle \psi | \hat H |\psi \rangle= \langle \Theta | H_{\text{eff}} | \Theta \rangle
\end{eqnarray}
where $\Theta = \Lambda^{[2]}B^{[2]}B^{[3]}$ is a tensor of order 4.  $H_{\text{eff}}$, which is the tensor of order 8, is obtained by the contraction of the left environment $L^{[2]}$, right environment $R^{[3]}$, $W^{[2]}$ and $W^{[3]}$.
Here the left  environment $L^{[i]}$  is obtained from the contraction of
 $L^{[i-1]}$, $W^{[i-1]}$, $A^{[i-1]}$ and $\overline{A^{[i-1]}}$, $L^{[0]}_{\beta_0\gamma_0 \alpha_0}= \delta_{\beta_0\gamma_0} v^L_{\alpha_0}$.  Similarly,
 $R^{[i]}$ is obtained from the contraction of
 $R^{[i+1]}$, $W^{[i+1]}$ and $B^{[i+1]}$, $\overline{B^{[i+1]}}$, $R^{[N]}_{\beta_N\gamma_N \alpha_N}=
  \delta_{\beta_N\gamma_N}v^R_{\alpha_N}$.

To keep the mixed canonical form of tensor of order $N$, The SVD for the minimum $\Theta_{\min}$ is
\begin{eqnarray}\label{23_7_5_22}
\Theta_{\min} =\tilde A^{[2]}\tilde \Lambda^{[3]}\tilde B^{[3]} \Longleftrightarrow
\Theta_{(\beta_2s_2^\prime),(\beta_4s_3^\prime)} =\sum_{\beta_3=1}^{\min(d\chi_2,d\chi_4)}\tilde A^{[2]}_{(\beta_2s_2^\prime),\beta_3}\tilde \Lambda^{[3]}_{\beta_3\beta_3}\tilde B^{[3]}_{\beta_3,(\beta_4s_3^\prime)}
\end{eqnarray}
Here the tensor $\Theta_{\min \beta_2\beta_4}^{s_2^\prime s_3^\prime}$ of order 4 is rewritten as a matrix
$\Theta_{(\beta_2s_2^\prime),(\beta_4s_3^\prime)}$. The bond dimension becomes $d$ times $\chi_2$ or $\chi_4$ which
is almost $d$ times of the bond dimension for the original $ \sum_{\beta_3=1}^{\chi_3}(A^{[2]s_2^\prime})_{\beta_2,\beta_3} \Lambda^{[3]}_{\beta_3\beta_3} (B^{[3]s_3^\prime})_{\beta_3,\beta_4}$. Here we assume that $\chi_2\sim \chi_3\sim \chi_4$.  To keep the bond dimension
under control, the SVD in (\ref{23_7_5_22}) is truncated for $k=1,\cdots, \chi_{\max}$. In practice, the largest
 $\chi_{\max}$ singular values in magnitude $\{\Lambda^{[3]}_{\beta_3\beta_3}\}_{\beta_3=1}^{\chi_{\max}}$ are kept
 and then scale these singular values such that $\sum_{k=1}^{\chi_{\max}} |\Lambda^{[3]}_{\beta_3\beta_3}|^2  =
 \sum_{k=1}^{\min(d\chi_2,d\chi_4)} |\tilde\Lambda^{[3]}_{\beta_3\beta_3}|^2 $ to ensure $|\psi_s|^2=1$.
For the ground state with not highly entangled, the singular values in magnitude decays vary fast. This truncation will not cause no much error.
Replacing $\Theta = \Lambda^{[2]}B^{[2]}B^{[3]}$ by
$\Theta_{\min} =\tilde A^{[2]}\tilde \Lambda^{[3]}\tilde B^{[3]}$, we get another better tensor with the less value of
 $\langle \psi | \hat O |\psi \rangle$. This better tensor is also in the mixed
canonical form. This is called one step of two-sites DMRG algorithm since the local tensors
$\Lambda^{[2]}B^{[2]}$ at site 2 and $B^{[3]}$ at site 3 are updated. To update the local tensors
$\Lambda^{[3]}B^{[3]}$ at site 3 and $B^{[4]}$  at site 4 in the next step of DMRG, we should prepare the
left environment $L^{[3]}$, right environment $R^{[4]}$. $L^{[3]}$ is the contraction of $L^{[2]}$,  $W^{[2]}$,
$\tilde A^{[2]}$ and $\overline{\tilde A^{[2]}}$. So we use $N-1$ steps of DMRG to update all $N$ sub-tensors from
the left to the right hand side. This is called one sweep of DMRG. The next sweep is implemented from the right to the left hand side.  This algorithm will stop if $\langle \psi |\hat H |\psi \rangle$ will not decrease.

\section{Boltzmann machine learning}\label{23.11.7.1}
The quantum expectation value of an operator $\hat O$ on a non-normalized state
$|\psi \rangle$ can be written as a classical expectation value over the distribution $p(s) = |\psi(s)|^2/
\langle  \psi |\psi \rangle$ using
\begin{eqnarray}\label{23_7_7_0}
\langle  \hat O\rangle \equiv \frac{\langle  \psi | \hat O|\psi \rangle}{\langle  \psi |\psi \rangle}  =\mathbb{E}[\tilde{O}] \equiv \sum_s p(s) \tilde{O}(s)
\end{eqnarray}
where
\begin{eqnarray}\label{23_7_7_1}
\tilde{O}(s)=\frac{\langle s|\hat{O}| \psi\rangle}{\langle s | \psi\rangle}=\sum_{s^{\prime}} \frac{\psi(s^{\prime})}{\psi(s)}\langle s|\hat{O}| s^{\prime}\rangle
\end{eqnarray}
If the matrix presentation $O$ of the operator $\hat O$ under the basis
$|s^\prime\rangle$ is sparse, there are only $O(n)$ basis $|s\rangle$ such that $\langle s|\hat{O}| s^{\prime}\rangle$ are nonzero. Eq. (\ref{23_7_7_0}) shows that $\langle \hat O\rangle$ can be calculated by Monte Carlo sampling of $\tilde O(s)$ with the probability $p(s)$.

We use neutral network to describe the wave function $\psi(s)$, which depends on real parameters $\theta=\{\theta_k\}_{k=1}^p$, and thus denoted by $\psi_\theta(s)$. Denote by $\delta\theta_k$ the change of the $k$th component $\theta_k$.
$\psi_{\theta+\delta\theta_k}(s)$ has the expansion
\begin{eqnarray}\label{23_5_11_0}
\psi_{\theta+\delta\theta_k}(s) =
\psi_{\theta}(s) + \delta\theta_k\frac{\partial\psi_{\theta}(s)}{\partial\theta_k}+ O(\delta\theta_k^2)
\end{eqnarray}
Define the local operator $\hat \Psi_k$ with matrix elements
\begin{eqnarray}\label{23_5_11_1}
\langle s|\hat \Psi_k|s^\prime\rangle = \delta_{s,s^\prime}\Psi_k(s)
\end{eqnarray}
where we introduced the log-derivative of $\psi_\theta(s)$
\begin{eqnarray}\label{23_5_11_2}
\Psi_k(s)=\frac{\partial\ln \psi_\theta(s)}{\partial \theta_k} = \frac{1}{\psi_\theta(s)}
\frac{\partial \psi_\theta(s)}{\partial \theta_k}
\end{eqnarray}
Here $\Psi_k(s)$ depends on $\theta$. We thus have
\begin{eqnarray}\label{23.9.21.0}
\langle s|\hat \Psi_k| \psi_\theta\rangle = \Psi_k(s)\psi_\theta(s) = \frac{\partial \psi_\theta(s)}{\partial \theta_k}
\end{eqnarray}
 From (\ref{23_5_11_0}), one has
\begin{eqnarray}\label{23_5_11_3}
|\psi_{\theta+\theta_k}\rangle = (1+\delta\theta_k \hat \Psi_k)| \psi_\theta\rangle
\end{eqnarray}
where we omitted $O(\delta\theta_k^2)$.

$|\psi_\theta\rangle$ is normalized to bep
\begin{eqnarray}\label{23_5_11_4}
|v_{0,\theta}\rangle = \frac{|\psi_\theta\rangle}{\sqrt{\langle\psi_\theta|\psi_\theta\rangle}}
\end{eqnarray}
Define
\begin{eqnarray}\label{23_5_11_5}
|v_{k,\theta}\rangle = (\hat \Psi_k - \langle \hat \Psi_k\rangle)|v_{0,\theta}\rangle, \quad k=1,\cdots,p
\end{eqnarray}
where
\begin{eqnarray}\label{23_5_11_6}
\langle \hat \Psi_k\rangle= \langle v_{0,\theta}|\hat \Psi_k|v_{0,\theta}\rangle = \frac{\langle \psi_{\theta}|
\hat \Psi_k|\psi_{\theta}\rangle}{\langle \psi_{\theta}|\psi_{\theta}\rangle}
\end{eqnarray}
Thus $|v_{k,\theta}\rangle$ and $|v_{0,\theta}\rangle$ are orthogonal, $k=1,\cdots,p$. But
$\langle v_{k,\theta}|v_{k^\prime,\theta}  \rangle\neq 0$ for $k,k^\prime=1,\cdots,p$.

The norm squared of $|\psi_{\theta+\delta\theta_k}\rangle$ is
\begin{eqnarray}\label{23_5_11_7}
\langle \psi_{\theta+\delta\theta_k}|\psi_{\theta+\delta\theta_k}\rangle &=&  \langle \psi_\theta |(1+\delta\theta_k \hat \Psi_k)^* |(1+\delta\theta_k\hat \Psi_k)| \psi_\theta\rangle \nonumber \\
& = & \langle \psi_{\theta}|\psi_{\theta}\rangle \Big[1+2 \text{Re} (\delta \theta_k\langle \hat \Psi_k\rangle) + O(\delta \theta_k^2)\Big]
\end{eqnarray}

One has from (\ref{23_5_11_3})(\ref{23_5_11_7})
\begin{eqnarray}\label{23_5_11_8}
|v_{0,\theta+\delta\theta_k}\rangle = \frac{|\psi_{\theta+\delta\theta_k}\rangle}{\sqrt{\langle \psi_{\theta+\delta\theta_k}|\psi_{\theta+\delta\theta_k}\rangle}}
&=&|v_{0,\theta}\rangle + \Big[\delta\theta_k\hat \Psi_k-\text{Re} (\delta\theta_k\langle \hat \Psi_k\rangle)\Big]|v_{0,\theta}\rangle
+ O(\delta\theta_k^2) \nonumber \\
& = & \Big[1+i\text{Im} (\delta\theta_k\langle \hat \Psi_k\rangle)\Big]|v_{0,\theta}\rangle +\delta\theta_k |v_{k,\theta}\rangle
+ O(\delta\theta_k^2) \quad (\text{by } (\ref{23_5_11_5}))\nonumber \\
& = & e^{i\delta\phi}\Big[|v_{0,\theta}\rangle  +\delta\theta_k |v_{k,\theta}\rangle \Big]+ O(\delta\theta_k^2)
\end{eqnarray}
where $\delta\phi = \text{Im} (\delta\theta_k\langle \hat \Psi_k\rangle)$.

Let $\hat H$ be the Hamiltonian operator,
\begin{eqnarray}\label{23_5_11_9}
&&\frac{\partial}{\partial\theta_k} \frac{\langle \psi_\theta|\hat
H|\psi_\theta\rangle}{\langle \psi_\theta|\psi_\theta\rangle} =
\frac{\partial  }{\partial\theta_k} \langle v_{0,\theta}|\hat
H|v_{0,\theta}\rangle \\
= \lim_{\delta\theta_k\rightarrow 0}\frac{\langle v_{0,\theta+\delta\theta_k} |\hat H|v_{0,\theta+\delta\theta_k}\rangle - \langle v_{0,\theta}|\hat H|v_{0,\theta}\rangle  }{\delta\theta_k} \nonumber \\
& = & \langle v_{k,\theta}|\hat H|v_{0,\theta} \rangle  +
 \langle v_{0,\theta}|\hat H|v_{k,\theta} \rangle   = 2\text{Re} \Big[\frac{\langle \psi_\theta|\hat H(\hat \Psi_k-\langle \hat \Psi_k\rangle)|\psi_\theta\rangle}{\langle \psi_\theta|\psi_\theta\rangle }\Big]  \nonumber \\
&=&  2\text{Re} \Big[\frac{\langle \psi_\theta|\hat H\hat \Psi_k|\psi_\theta\rangle}{\langle \psi_\theta|\psi_\theta\rangle }
 - \frac{\langle \psi_\theta|\hat H|\psi_\theta\rangle}{\langle \psi_\theta|\psi_\theta\rangle }
 \frac{\langle \psi_\theta|\hat \Psi_k|\psi_\theta\rangle}{\langle \psi_\theta|\psi_\theta\rangle }\Big]
\end{eqnarray}

The $k$th parameter $\theta_k$ is updated according to the gradient descent method
\begin{eqnarray}\label{23_5_11_13}
\theta_k \leftarrow \theta_k - \eta f_k
\end{eqnarray}
where $f_k =\frac{\partial E_\theta}{\partial\theta_k}$, $E_\theta =\frac{\langle \psi_\theta|\hat H|\psi_\theta\rangle}{\langle \psi_\theta|\psi_\theta\rangle}$, $\eta>0$ is the learning rate.

If each component $\theta_k$ has a change $\delta\theta_k$, the wave function $\psi_\theta(s)$ has changed to be $\psi_{\theta+\delta\theta}(s)$. Let
$ |v_{0,\theta+\delta\theta}\rangle = |\psi_{\theta+\delta\theta}\rangle/\sqrt{\langle \psi_{\theta+\delta\theta_k}|\psi_{\theta+\delta\theta_k}\rangle}$.
Equ. (\ref{23_5_11_8}) is generalized to be
\begin{eqnarray}\label{23_5_11_30}
|v_{0,\theta+\delta\theta}\rangle
 =  e^{i\delta\phi}\Big[|v_{0,\theta}\rangle  +\sum_{k=1}^p\delta\theta_k |v_{k,\theta}\rangle \Big]+ O(\delta\theta^2)
\end{eqnarray}
where $\delta\phi = \sum_{k=1}^p\text{Im} (\delta\theta_k\langle
\hat \Psi_k\rangle)$.

The distance between
$|v_{0,\theta}\rangle $ and $|v_{0,\theta+\delta\theta}\rangle$ is
\begin{eqnarray}\label{23_5_11_31}
\delta s^2 = \min_{\delta\alpha}|| e^{-i\delta\alpha}|v_{0,\theta+\delta\theta}\rangle -
|v_{0,\theta}\rangle ||^2
\end{eqnarray}
When $\delta\alpha = \delta\phi$, $\delta s^2$ is minimized to be
\begin{eqnarray}\label{23_5_11_33}
\delta s^2 = \sum_{1\leq k,k^\prime\leq p} \langle v_{k,\theta}| v_{k^\prime,\theta}\rangle \delta\theta_k\delta\theta_{k^\prime} +o(\delta\theta^2) =  \sum_{1\leq k,k^\prime\leq p}S_{k,k^\prime}\delta\theta_k\delta\theta_{k^\prime}+o(\delta\theta^2)
\end{eqnarray}
with
\begin{eqnarray}\label{23_5_11_35}
S_{k,k^\prime} = \frac{ \langle v_{k,\theta}| v_{k^\prime,\theta}\rangle +  \langle v_{k^\prime,\theta}| v_{k,\theta}\rangle }{2} = \text{Re}\Big(\langle v_{k,\theta}| v_{k^\prime,\theta}\rangle\Big)
 =  \text{Re}\Big(\langle \hat \Psi_k^*\hat \Psi_{k^\prime}\rangle - \langle \hat \Psi_k\rangle^*  \langle \Psi_{k^\prime}\rangle\Big)
\end{eqnarray}
Here we used
\begin{eqnarray}\label{23_5_11_36}
\langle v_{k,\theta}| v_{k^\prime,\theta}\rangle  =
\Big\langle  v_{0,\theta} \Big|(\hat \Psi_k - \langle \hat \Psi_k\rangle)^*  (\hat \Psi_{k^\prime} - \langle \hat \Psi_{k^\prime}\rangle)\Big|v_{0,\theta}\Big\rangle
\end{eqnarray}

Find $\delta\theta$ such that (1)
$\Delta E = E_{\theta+\delta\theta} - E_{\theta} $ is minimized; (2) the distance $\delta s^2$ between
$|v_{0,\theta}\rangle $ and $|v_{0,\theta+\delta\theta}\rangle$ is as small as possible. Minimize
\begin{eqnarray}\label{23_5_11_37}
\Delta E+\mu \delta s^2 = \sum_k\frac{\partial E_\theta}{\partial \theta_k}\delta\theta_k+\mu
 \sum_{k,k^\prime}S_{k,k^\prime}\delta\theta_k\delta\theta_{k^\prime} + o(\delta\theta^2)
\end{eqnarray}
to find that $\delta\theta$ satisfies
\begin{eqnarray}\label{23_5_11_36}
\sum_{k^\prime} S_{k,k^\prime} \delta\theta_{k^\prime}  =  -\frac{f_k}{2\mu}
\end{eqnarray}
which can be used to update the parameter $\theta$. This algorithm is called the Stochastic Reconfiguration method.

In the Boltzmann machine ansatz, the wave function is chosen to be
\begin{eqnarray}\label{23_5_11_54}
\psi_\theta(s)= \exp\Big(\sum_{j=0}^{N-1} a_j s_j\Big) \prod_{i=0}^{M-1} 2 \cosh \Big(b_i + \sum_{j=0}^{N-1} W_{ij}s_j\Big)
\end{eqnarray}
where $\theta = \{a_j, b_i, W_{ij}\}$ are the real parameters. Here $M$ is a given positive integer number.
$\alpha = M/N$ is the ratio of $M$ and $N$.
Comparing to the other neutral network ansatz, $\Psi_k$ for the Boltzmann machine ansatz can be calculated analytically \cite{Carleo_602}.

\section{Variational Quantum Eigensolver}\label{23.11.7.5}
The variational quantum eigensolver is aimed at finding the ground
state energy of a Hamiltonian $\hat H$ by minimizing the cost function $L(\theta) = \langle\psi_\theta|\hat H|\psi_\theta\rangle$, where he parameter $\theta$ is updated by $\theta  - \eta \nabla L(\theta)$ by iterations. Here
$\eta>0$ is the learning rate.
We choose the
trial state $|\psi_\theta\rangle = U_\theta|\psi_0\rangle$ for some ansatz $U_\theta $ and
initial state $|\psi_0\rangle$.
The variational Hamiltonian ansatz also aims to prepare a trial ground states
for a given Hamiltonian $\hat H = \sum_k \hat H_k$ (where $\hat H_k$ are Hermitian operators, usual Pauli strings) by Trotterizing an adiabatic state preparation process \cite{Wecker042303}. Here, each Trotter step corresponds to a variational ansatz so that the unitary is given $U_\theta = \prod_l (\prod_k e^{-i\theta_{l,k}\hat H_k})$.

Consider a general Hamiltonian
\begin{eqnarray}\label{23.9.21.10}
\hat H &=& \sum_{\sigma=\uparrow,\downarrow}\sum_{j=0}^{N-2} \Big(C_X^{\text{h}}  X_{j,\sigma} X_{j+1,\sigma} +
C_Y^{\text{h}} Y_{j, \sigma} Y_{j+1, \sigma}
+C_Z^{\text{h}} Z_{j, \sigma} Z_{j+1, \sigma}  \Big)   \nonumber \\
&& \hspace{0.5cm}  +  \sum_{j=0}^{N-1} \Big(
C_X^{\text{v}} X_{j, \uparrow} X_{j, \downarrow}+
C_Y^{\text{v}} Y_{j, \uparrow} Y_{j, \downarrow}+
C_Z^{\text{v}} Z_{j, \uparrow} Z_{j, \downarrow}\Big)
\end{eqnarray}
where $N$ is an even integer and the coefficients $C^{\text{h}}_\alpha>0$ and $C^{\text{v}}_\alpha$ does not depend on sites. $X_{j,\sigma}$, etc., denotes the Pauli $X$ matrix acting on site $j$ and spin $\sigma = \uparrow,\downarrow$.  Splitting the $N$ sites to be even sites and odd sites, this Hamiltonian can be written as
\begin{eqnarray}\label{23.9.21.11}
\hat H = \sum_{\alpha = X,Y,Z}  \Big( H^{\text{h,e}}_{\alpha} +  H^{\text{h,o}}_{\alpha} +
 H^{\text{v,e}}_{\alpha} +  H^{\text{v,o}}_{\alpha}    \Big) = H^{\text{e}}+H^{\text{o}}
\end{eqnarray}
where
\begin{eqnarray}\label{23.9.21.11.1}
  H^{\text{e}}  =  \sum_{\alpha =X,Y,Z} \Big( H^{\text{h,e}}_{\alpha} + H^{\text{v,e}}_{\alpha}\Big), \quad
  H^{\text{o}}  =  \sum_{\alpha =X,Y,Z} \Big( H^{\text{h,o}}_{\alpha}+H^{\text{v,o}}_{\alpha} \Big)
\end{eqnarray}
\begin{eqnarray}\label{23.9.21.12}
  H^{\text{h,e}}_{\alpha}  = \sum_{\sigma=\uparrow,\downarrow}\sum_{\text{even }j=0 }^{N-2} C_\alpha^{\text{h}} \alpha_{j, \sigma} \alpha_{j+1, \sigma}  , \quad  H^{\text{v,e}}_{\alpha}  = \sum_{\text{even }j=0 }^{N-1}  \alpha_{j, \uparrow} \alpha_{j, \downarrow} , \quad \alpha = X,Y,Z
\end{eqnarray}
$   H^{\text{h,o}}_{\alpha} $ and  $H^{\text{v,o}}_{\alpha}$ are defined similarly by replacing
summation over even site $j$ with odd site $j$.

Inspired by Ref. \cite{Wiersema_020319}\cite{Wen_2019},
the quantum circuit is $U_\theta = \prod_{k=1}^p U^{\text{h}}(\theta_k^{\text{h}})U^{\text{v}}(\theta_k^{\text{v}})$, where
\begin{eqnarray}\label{23.9.21.13}
U^{\text{h}}(\theta_k^{\text{h}})&=&
G(\theta^{\text{h,e}}_{k,X},H^{\text{h,e}}_{X})
G(\theta^{\text{h,e}}_{k,Y},H^{\text{h,e}}_{Y})
G(\theta^{\text{h,e}}_{k,Z},H^{\text{h,e}}_{Z}) \nonumber \\
&& G(\theta^{\text{h,o}}_{k,Z},H^{\text{h,o}}_{Z})
G(\theta^{\text{h,o}}_{k,Y},H^{\text{h,o}}_{Y})
G(\theta^{\text{h,o}}_{k,X},H^{\text{h,o}}_{X})
\end{eqnarray}
where $G(x,H) = e^{-ixH}$.
$U^{\text{v}}(\theta_k^{\text{v}})$ is defined similarly, with the subscription $\text{h}$ replaced by $\text{v}$.
Since the terms in $H^{\text{h,o}}_{\alpha}$, etc., commute with each other,
$$ G(\theta^{\text{h,e}}_{k,X},H^{\text{h,e}}_{X}) =
\prod_{\sigma=\uparrow,\downarrow}\prod_{\text{even }j=0 }^{N-2} e^{-i C_\alpha^{\text{h}} \alpha_{j, \sigma} \alpha_{j+1, \sigma}} $$

The initial state $|\psi_0 \rangle$ is chosen to be  the  ground state
 $|\psi_0 \rangle = \otimes_{\sigma=\uparrow,\downarrow}\otimes_{\text{even }j=0}^{N-2} \frac{1}{\sqrt{2}}(|01\rangle - |10\rangle)$ of $\sum_{\alpha =X,Y,Z} H^{\text{h,e}}_{\alpha}$
 with the ground state energy $-3NC_\alpha^{\text{h}}$. This is because
\begin{eqnarray}\label{23.9.22.1}
\alpha_{j, \sigma} \alpha_{j+1, \sigma} (|01\rangle - |10\rangle) = -
(|01\rangle - |10\rangle)
\end{eqnarray}
for $\alpha = X,Y,Z$, $\sigma=\uparrow,\downarrow$ and even site $j=0, 2, \cdots, N-2$.

\section{Fermi-Hubbard model}\label{23.9.19.1}

The Hamiltonian for Fermi-Hubbard model is
\begin{eqnarray}\label{23_5_21_1}
\hat H &=& - t\sum_{\sigma=\uparrow,\downarrow}\sum_{j=0}^{N-2}  (c^{\dagger}_{j, \sigma} c_{j+1, \sigma} + \text{h.c.})
    +  U\sum_{j=0}^{N-1} n_{j, \uparrow} n_{j, \downarrow}
    - \mu\sum_{\sigma=\uparrow,\downarrow}\sum_{j=0}^{N-1}   n_{j, \sigma} \nonumber \\
 &=& - t\sum_{\sigma=\uparrow,\downarrow}\sum_{j=0}^{N-2}  (c^{\dagger}_{j, \sigma} c_{j+1, \sigma} + \text{h.c.})
    +  U\sum_{j=0}^{N-1} \Big(n_{j, \uparrow}-\frac{1}{2}\Big) \Big( n_{j, \downarrow}-\frac{1}{2}\Big)
    - \Big(\mu-\frac{U}{2}\Big)\sum_{\sigma=\uparrow,\downarrow}\sum_{j=0}^{N-1}   n_{j, \sigma} - \frac{NU}{4}
\end{eqnarray}
where open boundary condition is assumed and $n_{j,\sigma} = c^\dagger_{j,\sigma}c_{j,\sigma}$. The creation operator $c^\dagger_{j,\sigma}$ and annihilate operators $c_{j,\sigma}$ satisfy
\begin{eqnarray}\label{23_7_7_20}
 \{c_{j,\sigma} , c^\dagger_{k,\tau}\} = \delta_{jk}\delta_{\sigma\tau}, \quad \{c_{j,\sigma} , c_{k,\tau}\} = 0, \quad
\{c_{j,\sigma}^\dagger ,c^\dagger_{k,\tau}\} = 0
\end{eqnarray}
For the half filling case: $\mu = U/2$, the Hamiltonian is reduced to
\begin{eqnarray}\label{23_5_23_5}
\hat H       = - t\sum_{\sigma=\uparrow,\downarrow}\sum_{j=0}^{N-2}  (c^{\dagger}_{j, \sigma} c_{j+1, \sigma} + \text{h.c.})
    +  U\sum_{j=0}^{N-1} \Big(n_{j, \uparrow}-\frac{1}{2}\Big) \Big( n_{j, \downarrow}-\frac{1}{2}\Big)
       - \frac{NU}{4}
\end{eqnarray}
Combining the index $\sigma = \uparrow (0), \downarrow (1) $ and $j=0,\cdots,N-1$ to be $k= N\sigma + j $, we have  \begin{eqnarray}\label{23_5_23_5_1}
\hat H       = - \sum_{k=0}^{2N-2} t_k  (c^{\dagger}_{k} c_{k+1} + \text{h.c.})
    +  U\sum_{k=0}^{N-1} \Big(n_{k}-\frac{1}{2}\Big) \Big( n_{k+N}-\frac{1}{2}\Big)
       - \frac{NU}{4}
\end{eqnarray}
where $t_k=t$ if $k\neq N-1$ and $0$ if $k=N-1$.
The Jordan-Wigner representation shows that
$$ n_{j} = \frac{1}{2} + \frac{1}{2}Z_{j} \Longleftrightarrow
Z_{j}  =  2 n_{j} -1   $$
\begin{eqnarray*}
 2( c^{\dagger}_{k+1}c_{k} +   c^{\dagger}_{k}c_{k+1})
= X_{k}X_{k+1}  +Y_{k}Y_{k+1}
\end{eqnarray*}
Here $X, Y , Z$ are Pauli operators. In the Jordan-Wigner representation the Hamiltonian for half-filling case,
 is written as
\begin{eqnarray}\label{23_5_27_1}
 \frac{1}{tN} \hat H =
-\frac{1}{2N}\sum_{k=0}^{2N-2} t_k  (X_{k} X_{k+1} +
Y_{k} Y_{k+1} )
    + \frac{U}{4tN} \sum_{k=0}^{N-1} Z_{k} Z_{k+N}
    -\frac{U}{4t}
\end{eqnarray}
which is the special case for the Hamiltonian in (\ref{23.9.21.10}) except the constant $-\frac{U}{4t}$.

If one use the index $k=2j+\sigma$ for $j=0,\cdots,N-1$ and $\sigma= \uparrow (0), \downarrow (1) $, the Hamiltonian in (\ref{23_5_23_5}) becomes
\begin{eqnarray}\label{23.11.15.1}
\hat H  = - t  \sum_{k=0}^{2N-3} ( c^{\dagger}_{k} c_{k+2} + \text{h.c.} )
    +  U\sum_{k=0}^{N-1} \Big(n_{2k}-\frac{1}{2}\Big) \Big( n_{2k+1}-\frac{1}{2}\Big)
       - \frac{NU}{4}
\end{eqnarray}
Under the Jordan-Wigner representation
\begin{eqnarray*}
 2( c^{\dagger}_{k+2}c_{k} +   c^{\dagger}_{k}c_{k+2})
= X_{k}(-Z_{k+1})X_{k+2}  +Y_{k}(-Z_{k+1})Y_{k+2}
\end{eqnarray*}
it becomes
\begin{eqnarray}\label{23.11.15.2}
\frac{1}{tN}\hat H  = - \frac{1}{2N}  \sum_{k=0}^{2N-3} \Big( X_k (-Z_{k+1}) X_{k+2} + Y_k (-Z_{k+1}) Y_{k+2} \Big)
    + \frac{ U}{4t}\sum_{k=0}^{N-1} Z_{2k} Z_{2k+1}
       - \frac{U}{4t}
\end{eqnarray}

\section{Schwinger model}\label{23.11.5.0}
The massive Schwinger model, or QED in two space-time dimensions, is a gauge
theory describing the interaction of a single fermionic species with the photon field. In the temporal gauge,
the Hamiltonian density is
\begin{eqnarray}\label{2021_7_9_1}
\mathcal{H}=-i \psi^{\dagger} \gamma^{0} \gamma^{1}\left(\partial_x+i g A\right) \psi+m \psi^{\dagger}(x) \gamma^{0} \psi+\frac{1}{2} E^{2}
\end{eqnarray}
where $$\gamma^{0}=\left(\begin{array}{cc}
1 & 0 \\
0 & -1
\end{array}\right),  \quad \gamma^{1}=\left(\begin{array}{cc}
0 & 1 \\
-1 & 0
\end{array}\right)
$$ are the $\gamma$ matrices,
$m$ and $g$ are two parameters. The electric field, $E= \partial_t A$ satisfies the Gauss's law
$$\partial_x E(x) - g \psi^\dagger(x) \psi(x) = 0$$
Here $\partial_x$ and $\partial_t$ denotes the partial derivative along one dimensional space direction $x$ and time direction $t$.

The theory is quantized using canonical quantization by imposing anticommutation relations on the fermion fields
\begin{eqnarray}\label{23.11.5.10}
\left\{\psi^{\dagger}(x, t), \psi(y, t)\right\}=\delta(x-y)
\end{eqnarray}
and by imposing commutation relations on the gauge fields
\begin{eqnarray}\label{23.11.5.11}
\left[E(x, t), A(y, t)\right]=i \delta(x-y)
\end{eqnarray}

The model can now be formulated on a ``staggered" spatial lattice. Let the lattice spacing be $a$, and label the sites with an integer $n$. Define a single-component fermion field $\phi(n)$ at each site $n$, obeying anticommutation relations
\begin{eqnarray}\label{2021_7_9_3}
\left\{\phi^{\dagger}(n), \phi(m)\right\}=\delta_{m n}
\end{eqnarray}
The gauge field is defined on the links $(n, n+1)$ connecting each pair of sites by
\begin{eqnarray}\label{2021_7_9_4}
U(n, n+1)=e^{i \theta(n)}=e^{i a g A(n)}
\end{eqnarray}
Then the lattice Hamiltonian equivalent to Eq. (\ref{2021_7_9_1}) is
\begin{eqnarray}\label{2021_7_9_5}
\hat H=\frac{-i}{2 a} \sum_{n=0}^{N-1}\left[\phi^{\dagger}(n) e^{i \theta(n)} \phi(n+1)-\text {H.c.}\right]
+m \sum_{n=0}^{N-1}(-1)^{n} \phi^{\dagger}(n) \phi(n)+\frac{g^{2} a}{2} \sum_{n=0}^{N-1} L^{2}(n)
\end{eqnarray}
where the number of lattice sites $N$ is even, and the correspondence between lattice and continuum fields is
\begin{eqnarray}\label{2021_7_9_6}
\frac{\phi(n)}{\sqrt{a}} \rightarrow \begin{cases}\psi_e(x), & n \text { even, } \\ \psi_o(x), & n \text { odd }\end{cases}
\end{eqnarray}
$\psi_e$ and $\psi_o$ are the two components of the continuum spinor $\psi = (\psi_e,\psi_o)$,
\begin{eqnarray}\label{2021_7_9_7}
\frac{1}{a g} \theta(n) \rightarrow A(x), \quad
g L(n) \rightarrow E(x)
\end{eqnarray}
From the commutation relations (\ref{23.11.5.11}) and the above correspondence, one has
\begin{eqnarray}\label{2021_7_9_8}
[\theta(n), L(m)]=i \delta_{n m}
\end{eqnarray}
The dimensionless Hamiltonian is $\frac{2}{a g^{2}} \hat H$, which is also denoted by $\hat H$
\begin{eqnarray}\label{21_8_17_0}
\hat H=-i x \sum_{n=0}^{N-1}\left[\phi^{\dagger}(n) e^{i \theta(n)} \phi(n+1)-\text {H.c.}\right]
+\mu \sum_{n=0}^{N-1}(-1)^{n} \phi^{\dagger}(n) \phi(n)+ \sum_{n=0}^{N-1} L^{2}(n)
\end{eqnarray}
with the dimensionless parameters
\begin{eqnarray}\label{2021_7_9_9}
x=\frac{1}{g^{2} a^{2}}, \quad \mu=\frac{2 m}{g^{2} a}
\end{eqnarray}
The Gauss law $\partial_{x} E  = g \psi^\dagger  \psi $ is discretized to be
\begin{eqnarray}\label{2021_7_9_9_0}
L(n)-L(n-1)=\phi(n)^{\dagger} \phi(n)-\frac{1}{2}\left[1-(-1)^{n}\right]
\end{eqnarray}

The elimination of fermion fields can be realized by using a Jordan-Wigner transformation
\begin{eqnarray*}
\phi(n)=\prod_{k<n}\left(- \sigma_{k}^{z}\right) \sigma_{n}^{-}
, \quad \phi(n)^\dagger=\prod_{k<n}\left(- \sigma_{k}^{z}\right)\sigma_{n}^{+}
\end{eqnarray*}
where $\sigma_n^+ = (X_n+iY_n)/2$ ($\sigma_n^-= (X_n-iY_n)/2$) denotes the matrix $|0\rangle \langle 1|$ ($|1\rangle \langle 0|$) acting on the $j$th qubit.

The Gauss law in (\ref{2021_7_9_9_0}) is reduced to be
\begin{eqnarray}\label{2021_7_9_9_0_0}
L(n)-L(n-1)= \frac{1}{2} [ \sigma_{n}^{z} + (-1)^{n}]
\end{eqnarray}
and the dimensionless Hamiltonian in (\ref{21_8_17_0}) is
\begin{eqnarray}\label{21_8_17_0_0}
H=-i x \sum_{n=0}^{N-1}\left[\sigma_{n}^{+}e^{i \theta(n)}\sigma_{n+1}^{-}
-\text {H.c.}\right]
+\mu \sum_{n=0}^{N-1}(-1)^{n}  \frac{1}{2}(\sigma_{n}^{z} + 1)+ \sum_{n=0}^{N-1} L^{2}(n)
\end{eqnarray}
The factor $e^{i \theta(n)}$ has been eliminated by a residual gauge transformation
\cite{Hamer_9701015}
\begin{eqnarray}\label{21_8_16_2}
\sigma^{-}_n \rightarrow \prod_{l<n}\left\{e^{-i \theta(l)}\right\} \sigma^{-}_n
\end{eqnarray}

Finally, the model (\ref{21_8_17_0}) can be mapped to a spin 1/2 Hamiltonian \cite{Hamer_9701015}
\cite{Banuls_158}
\begin{eqnarray}\label{21_8_16_1}
H&=& x \sum_{n=0}^{N-2}\left[\sigma_{n}^{+} \sigma_{n+1}^{-}+\sigma_{n}^{-} \sigma_{n+1}^{+}\right]+\frac{\mu}{2}
 \sum_{n=0}^{N-1}\left[1+(-1)^{n} \sigma_{n}^{z}\right]
+\sum_{n=0}^{N-2}\left[\ell+\frac{1}{2} \sum_{k=0}^{n}\left((-1)^{k}+\sigma_{k}^{z}\right)\right]^{2} \nonumber \\
&=& \frac{x}{2} \sum_{n=0}^{N-2}\left[X_{n} X_{n+1}+Y_{n}Y_{n+1}\right]+\frac{\mu}{2}
 \sum_{n=0}^{N-1}\left[1+(-1)^{n} Z_{n}\right]
+\sum_{n=0}^{N-2}\left[\ell+\frac{1}{2} \sum_{k=0}^{n}\left((-1)^{k}+Z_{k}\right)\right]^{2}
\end{eqnarray}
where $\ell$ is the boundary electric field (on the leftmost link), which can describe the background field.
In the second equality we used
$$ \sigma_n^{\pm} = \frac{X_n\pm iY_n}{2}, \quad
\sigma_{n}^{+} \sigma_{n+1}^{-}+\sigma_{n}^{-} \sigma_{n+1}^{+} = \frac{X_{n} X_{n+1}+Y_{n}Y_{n+1}}{2}, \quad \sigma_n^z = Z_n $$

\section{Simulation results}\label{23.9.19.2}

\begin{figure}[h]
\centering
\includegraphics[width=8cm,height=6cm]{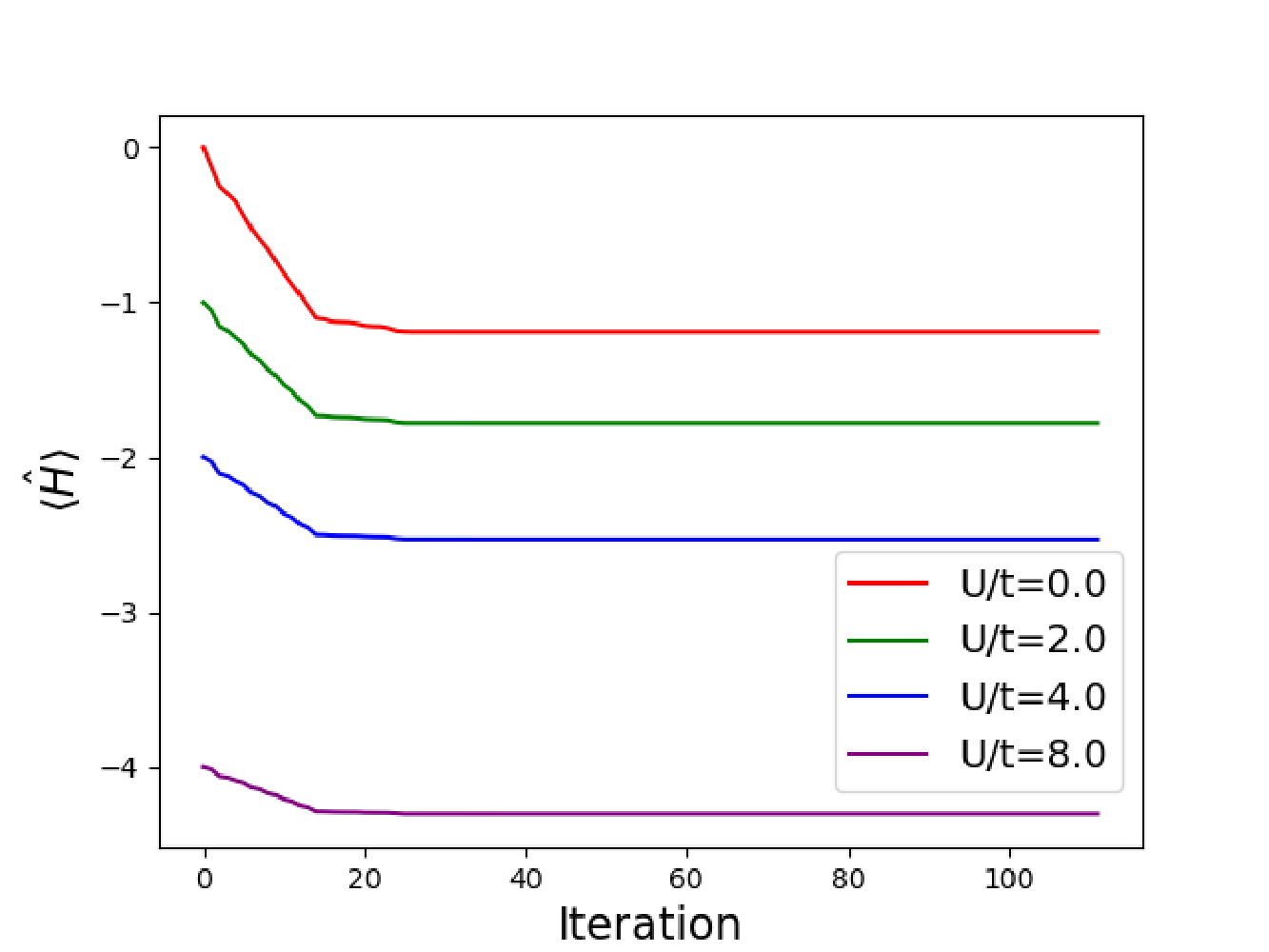}
\caption{The dependence of $\langle \hat{H}\rangle$ of the Fermi-Hubbard model for half-filling case on the iteration for different $U/t$, $N=8$ by DMRG algorithm.}\label{23_9_25_0}
\end{figure}

We investigate the Fermi-Hubbard model at half filling, where the Hamiltonian is given by (\ref{23_5_23_5}) or (\ref{23_5_27_1}). Table \ref{23_5_24_10} presents the exact ground state energy per site $E_0/N$ obtained by exact diagonalization for different values of $N$ and $U/t$. It is evident that the thermodynamic limit is achieved in each column. Therefore, we choose the largest system size, $N=8$, to study the problem.
Using the Hamiltonian (\ref{23_5_27_1}), we employ the DMRG algorithm to calculate the ground state energy per site $E_0/N$. The results are displayed in Table \ref{23_5_27_3}, where we compare them with the exact values. The maximum error between the DMRG and exact ground state energies is approximately $1.413783\times 10^{-9}$ for $(U/t,N)=(0,8)$, demonstrating the high accuracy of the DMRG method.
Figure \ref{23_9_25_0} illustrates the energy $\langle\hat H \rangle=\langle \psi | \hat H |\psi \rangle$ in equation (\ref{23_7_5_8}) as a function of iteration for different values of $U/t$ with $N=8$. We perform 8 sweeps, with 14 ($=2(N-1)$) local minimizations in each sweep (multiplied by 2 due to spin up and spin down states). It is noteworthy that the ground state energy is obtained after a single sweep (14 local minimizations). Furthermore, the energy $\langle\hat H \rangle$ decreases with each step of local minimization.

\begin{figure}[h]
\centering
\includegraphics[width=8cm,height=6cm]{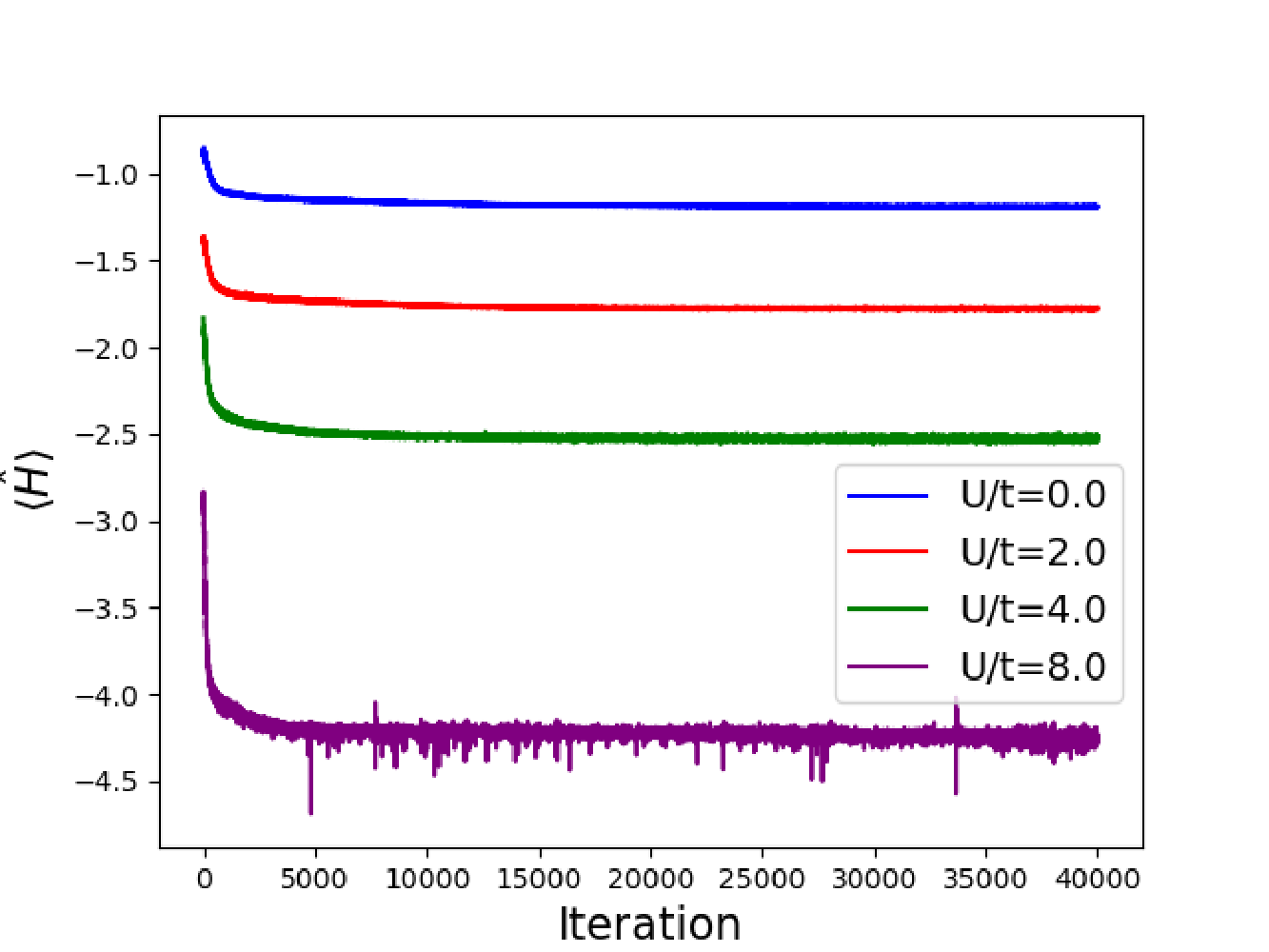}
\caption{The dependence of $\langle \hat{H}\rangle$  of the Fermi-Hubbard model for half-filling case on the iteration for different $U/t$ by restricted Boltzmann machine with $\alpha=4$ and $N=8$.}\label{23_9_26_1}
\end{figure}

Table \ref{23.10.26.0} displays the ground state energy per site, $E_0/N$, obtained using the restricted Boltzmann machine with $\alpha=4$. To calculate the ground state energy $E_0/N$, we use the last 10,000 sampling data out of a total of 40,000 data points. Comparing these results with those obtained using the DMRG algorithm, we find that the results from the restricted Boltzmann machine are reliable when $U/t$ and $N$ are not too large. However, for $U/t=8$ and $N=8$, the value $E_0/N=-4.211473$ in Table \ref{23.10.26.0} is significantly higher than $E_0/N=-4.3026039$ in Table \ref{23_5_27_3}. Figure \ref{23_9_26_1} illustrates the dependence of $\langle\hat H\rangle$ on the iteration number for the restricted Boltzmann machine with $\alpha=4$ and $N=8$. For the case $U/t=8.0$, there are large fluctuations of $\langle\hat H\rangle$ with each iteration.

\begin{figure}[h]
\centering
\includegraphics[width=8cm,height=6cm]{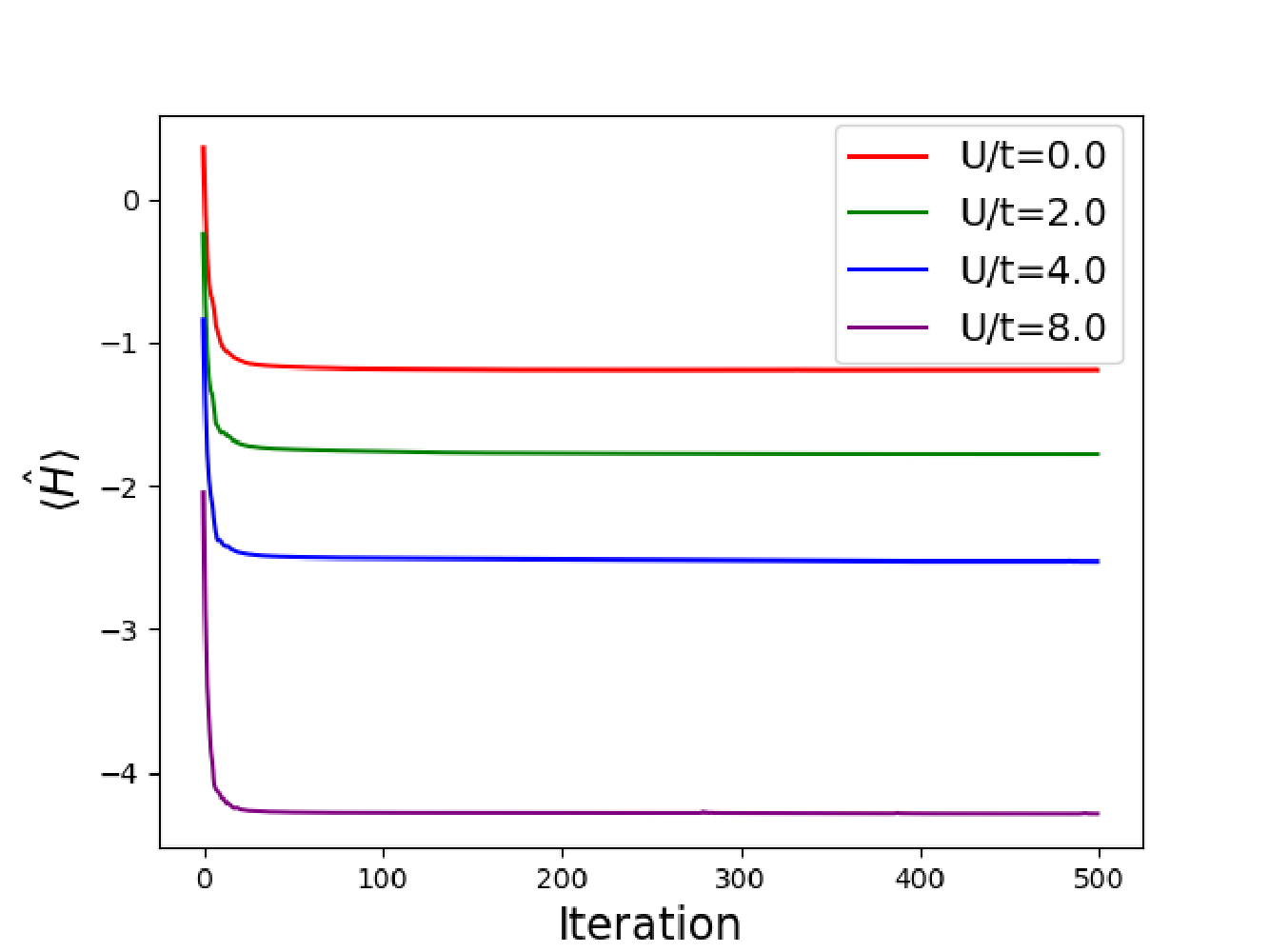}
\caption{The dependence of $\langle \hat{H}\rangle$  of the Fermi-Hubbard model for half-filling case on the iteration by VQE with $p=6$ for different $U/t$, $N=8$.}\label{23_9_26_0}
\end{figure}

Table \ref{23_5_24_10_1} displays the ground state energy per site, $E_0/N$, obtained using the variational quantum eigensolver (VQE) with $p=6$. The results from VQE show good agreement with those obtained using the DMRG algorithm. The largest deviation in $E_0/N$ between VQE and DMRG is 0.0176116, which occurs for the case $U/t=8$ and $N=8$.
Figure \ref{23_9_26_0} illustrates the dependence of $\langle\hat H\rangle$ on the iteration number for VQE with different values of $U/t$ and $N=8$. When compared with the fluctuation in $\langle\hat H\rangle$ shown in Figure \ref{23_9_26_1} for the restricted Boltzmann machine, the fluctuation observed in Figure \ref{23_9_26_0} is very small.

The ground state energy per site, $E_0/N$, of the Schwinger model (\ref{21_8_16_1}) has been calculated using both the density matrix renormalization group (DMRG) algorithm and the variational quantum eigensolver (VQE). However, the restricted Boltzmann machine is not suitable for solving the Schwinger model as it fails to approach the ground state energy accurately. In this study, we have chosen $x=100$ and $\ell=0$. p
Table \ref{23.11.5.20} presents the ground state energy per site obtained by the DMRG algorithm for different values of $\mu$ and $N$. Comparing the results to the exact values, we observe that $E_0/N$ for $N=4$ and $8$ has an error less than $O(10^{-10})$, while for $N=16$ the error is on the order of $O(10^{-4})$. For fixed values of $\mu=0$ and $0.5$, $E_0/N$ tends to the thermodynamic limit as $N\rightarrow \infty$.
Table \ref{23.11.5.21} displays the ground state energy per site obtained by VQE for different values of $\mu$ and $N=4, 8$, and $16$ due to computational limitations. It is worth noting that $E_0/N$ obtained by VQE is slightly larger than those obtained by the DMRG algorithm.

\section{Discussion}\label{23.11.8.0}
We have used three variational methods to calculate the ground state energy for Fermi-Hubbard model at half filling and Schwinger model.  The DMRG algorithm has proven to be successful for these two models, while the VQE algorithm, although less accurate, is also effective for calculating ground state energies. However, it seems that the restricted Boltzmann learning method is not as accurate as the DMRG algorithm for the Fermi-Hubbard model. It's unfortunate that the restricted Boltzmann does not work for solving the ground state energy of Schwinger model. We also used mean field ansatz, Jastrow ansatz, multi-layer dense network, autoregressive networks \cite{Sharir_124} and group convolutional neural networks \cite{Roth_05085}\cite{Cohen_07576}, etc., but it also does not work.

Variational methods can indeed be applied to various other aspects
of quantum systems, such as excited states, real-time dynamics, and
dissipative dynamics. It's beneficial to learn from algorithms and
ideas used in one variational method and apply them to other
variational methods. This cross-pollination of ideas can lead to
further
 advancements in the field.

\begin{table}[h]
\centering
\begin{tabular}{| c | c | c | c | c |  }
\hline
$U/t$  & 0 & 2 & 4 & 8 \\
\hline
$N=4$  & -1.11803398874989 & -1.71898570225126 & -2.48828632717113  & -4.27929310334025 \\
$N=6$  & -1.16465306914497 & -1.75771896573933 & -2.51542755325090 &  -4.29468312587687 \\
$N=8$  & -1.18969262078591 & -1.77820426808516 & -2.52947587489120 & -4.30260392500470 \\
\hline
\end{tabular}
\caption{The ground state energy $E_0/N$ obtained by exact diagonalization for different $U/t$ and $N$.}\label{23_5_24_10}
\end{table}

\begin{table}[h]
\centering
\begin{tabular}{| c | c | c | c | c |  }
\hline
$U/t$  & 0 & 2 & 4 & 8 \\
\hline
$N=4$ &  -1.11803398874989 & -1.71898570225126 & -2.48828632717113 & -4.2792931033402    \\
$N=6$ & -1.16465306914497 & -1.75771896573933 & -2.51542755325092 & -4.2946831258768   \\
$N=8$ & -1.18969261937212 & -1.77820426730531 & -2.52947587474444  & -4.3026039250042   \\
\hline
\end{tabular}
\caption{The ground state energy $E_0/N$ of the Fermi-Hubbard model for half-filling case obtained by DMRG algorithm.}\label{23_5_27_3}
\end{table}

\begin{table}[h]
\centering
\begin{tabular}{| c | c | c | c | c |  }
\hline
$U/t$  & 0 & 2 & 4 & 8 \\
\hline
$N=4$  & -1.112495 $\pm$ 0.00458 & -1.713604 $\pm$ 0.00430 & -2.482209 $\pm$ 0.00468  & -4.270301 $\pm$ 0.00639 \\
$N=6$  & -1.154108 $\pm$ 0.00698 & -1.747218 $\pm$ 0.00706 &-2.502630 $\pm$ 0.00775  & -4.272329 $\pm$ 0.01133 \\
$N=8$  & -1.172597 $\pm$ 0.00820 & -1.759270 $\pm$ 0.00934 & -2.507544 $\pm$ 0.01090 & -4.211473 $\pm$ 0.01207 \\
\hline
\end{tabular}
\caption{The ground state energy $E_0/N$  of the Fermi-Hubbard model for half-filling case obtained from the last 10000 sampling data (total 40000 data). Using restricted Boltzmann machine with $\alpha=4$.} \label{23.10.26.0}
\end{table}

\begin{table}[h]
\centering
\begin{tabular}{| c | c | c | c | c |  }
\hline
$U/t$  & 0 & 2 & 4 & 8 \\
\hline
$N=4$  & -1.1180205962 $\pm$ 0.0000029564  & -1.7188572935 $\pm$ 0.0000369740  & -2.4882732507 $\pm$ 0.0000091629  & -4.2743820234 $\pm$ 0.0023639140 \\
$N=6$  &  -1.1646497137 $\pm$ 0.0000005722 & -1.7574496710 $\pm$ 0.0000149607 & -2.5146574899 $\pm$ 0.0000709233  & -4.2844134948 $\pm$ 0.0007170862 \\
$N=8$  & -1.1896184621 $\pm$ 0.0000047168 & -1.7765344066 $\pm$ 0.0001591077 & -2.5201447110 $\pm$ 0.0012651519 & -4.2849922854 $\pm$ 0.0013377152\\
\hline
\end{tabular}
\caption{The ground state energy $E_0/N$  of the Fermi-Hubbard model for half-filling case obtained by VQE after 2000 iterations, $p=6$ in quantum circuit $U(\theta)$, the learning rate $\eta=0.01$.}\label{23_5_24_10_1}
\end{table}

\begin{table}[h]
\centering
\begin{tabular}{| c | c | c | c | c |  }
\hline
$\mu$  & 0.0 & 2.5 \\
\hline
$N=4$ &  $-55.6279081848$ & $-54.4037370017$ \\
\hline
$N=8$ &  $-59.1849689348$ & $-57.9669117540$ \\
\hline
$N=16$ & $-61.1628081594$ & $-59.9540535908$ \\
\hline
$N=32$ & $-62.2090479768$ & $-61.0124235459$ \\
\hline
$N=64$ & $-62.7525150696$ & $-61.5643604631$ \\
\hline
\end{tabular}
\caption{The ground state energy $E_0/N$ of the Schwinger model with $x=100$ obtained by DMRG algorithm.} \label{23.11.5.20}
\end{table}

\begin{table}[h]
\centering
\begin{tabular}{| c | c | c | c | c |  }
\hline
$\mu$  & 0.0 & 2.5 \\
\hline
$N=4$  &  -55.5133966685 $\pm$ 0.0004353889 & -54.2636722061  $\pm$ 0.0002959286 \\
$N=8$  &  -58.8314466647 $\pm$ 0.0005823967 & -57.5816064928  $\pm$ 0.0004639107\\
$N=16$  & -60.3183145037 $\pm$ 0.0005032675 & -59.0682375601  $\pm$ 0.0004984761 \\
\hline
\end{tabular}
\caption{The ground state energy $E_0/N$ of the Schwinger model with $x=100$ obtained by VQE. The last 1000 iterations are used (total 3000 iterations), $p=6$ in quantum circuit $U(\theta)$, and the learning rate $\eta=0.01$.} \label{23.11.5.21}
\end{table}

\appendix

\section{\label{Appendix_0} The decompsiton in (\ref{22_2_5_17})}
For the sake of notation convenience, $s_i=1,\cdots,d$ for $i=0,\cdots,N-1$.
Combining the last $N-1$ indices $\{s_i\}_{i=1}^{N-1}$ together:
$(s_{1}\cdots s_{N-1}) = (s_{1}-1)d^{N-2} + \cdots + (s_{N-1}-1)d^0+1\leq d^{N-1}$,
the tensor $\psi_{s_{0},\cdots,s_{N-1}}$ can be regarded as $d\times d^{N-1}$ matrix
\begin{eqnarray}\label{22_2_5_23}
\psi_{s_{0},\cdots,s_{N-1}} \rightarrow \psi_{s_{0},(s_{1}\cdots s_{N-1})}
\end{eqnarray}

The SVD of $d\times d^{N-1}$ matrix $\psi_{s_0,(s_{1}\cdots s_{N-1})}$ is
\begin{eqnarray}\label{22_2_5_24}
    \psi_{s_0,(s_{1}\cdots s_{N-1})} = \sum_{\alpha_1 = 1}^{\chi_1}U_{s_0,\alpha_1}S_{\alpha_1} V_{\alpha_1,(s_{1}\cdots s_{N-1})}^{\dagger }
\end{eqnarray}
with $\chi_1\leq d$, $S_{\alpha_1}$ is the singular value, $U$ and $V$ are $d\times \chi_1$ and $d^{N-1}\times \chi_1$ unitary matrix (the column vectors with unit length are orthogonal to each other), respectively.
Let $(M^{[0]s_0})_{\alpha_0,\alpha_1} = U_{s_{0},\alpha_1}$, which is a $1 \times \chi_1$ matrix ($\alpha_0=1$ and $s_0$ fixed). The left two parts in SVD are combined to be $(\chi_1d)\times d^{N-2}$ matrix
\begin{eqnarray}\label{22_2_5_25}
        S_{\alpha_1}V_{\alpha_1,(s_{1}\cdots s_{N-1})}^{\dagger }
 \rightarrow \psi_{(\alpha_1s_{1}),(s_{2}\cdots s_{N-1})}
 \end{eqnarray}
where $(\alpha_1s_{1}) = (\alpha_1-1)d + s_{1}\leq (\chi_1-1)d+d = \chi_1 d$, $(s_{2}\cdots s_{N-1}) = (s_{2}-1)d^{N-3} + \cdots + (s_{N-1}-1)d^0+1\leq d^{N-2}$. So, we have
$$  \psi_{s_0,(s_{1}\cdots s_{N-1})} = \sum_{\alpha_1 = 1}^{\chi_1} (M^{[0]s_0})_{\alpha_0,\alpha_1}
\psi_{(\alpha_1s_{1}),(s_{2}\cdots s_{N-1})} $$

The SVD of $\psi_{(\alpha_1s_{1}),(s_{2}\cdots s_{N-1})}$ is
\begin{eqnarray}\label{22_2_5_26}
    \psi_{(\alpha_1s_{1}),(s_{2}\cdots s_{N-1})} = \sum_{\alpha_2 = 1}^{\chi_2}U_{(\alpha_1s_{1}),\alpha_2}S_{\alpha_2} V_{\alpha_2,(s_{2}\cdots s_{N-1})}^{\dagger }
\end{eqnarray}
Let $(M^{[1]s_1})_{\alpha_1,\alpha_2} = U_{(\alpha_1s_{1}),\alpha_2}$, which is a $\chi_1 \times \chi_2$ matrix ($s_1$ fixed). The remaining two parts are combined as  $(\chi_2d)\times d^{N-3}$ matrix
\begin{eqnarray}\label{22_2_5_25_1}
        S_{\alpha_2}V_{\alpha_2,(s_{2}\cdots s_{N-1})}^{\dagger }
 \rightarrow \psi_{(\alpha_2s_{2}),(s_{3}\cdots s_{N-1})}
 \end{eqnarray}

 After $N-2$ steps of SVD,  we get $(\chi_{N-2}d)\times d$ matrix $\psi_{(\alpha_{N-2}s_{N-2}),s_{N-1}}$, and use SVD again
 \begin{eqnarray}\label{22_2_5_27}
    \psi_{(\alpha_{N-2}s_{N-2}),s_{N-1}} = \sum_{\alpha_{N-1} = 1}^{\chi_{N-1}}U_{(\alpha_{N-2}s_{N-2}),\alpha_{N-1}}S_{\alpha_{N-1}} V_{\alpha_{N-1},s_{N-1}}^{\dagger }
\end{eqnarray}
Let $(M^{[N-2]s_{N-2}})_{\alpha_{N-2},\alpha_{N-1}} =
U_{(\alpha_{N-2}s_{N-2}),\alpha_{N-1}}$, which is a $\chi_{N-2}
\times \chi_{N-1}$ matrix. Let
$(M^{[N-1]s_{N-1}})_{\alpha_{N-1},\alpha_{N}} = S_{\alpha_{N-1}}
V_{\alpha_{N-1},s_{N-1}}^{\dagger }$, which is a $\chi_{N-1} \times
1$ matrix ($\alpha_{N}=1$ and $s_{N-1}=1$ fixed). Thus after $N-1$
steps of SVD, the decomposition in (\ref{22_2_5_17}) is obtained.

From the above SVD processes, $\{M^{[i]}\}_{i=0}^{N-2}$ satisfy the left canonical condition. For example,
\begin{eqnarray}\label{22_2_5_30}
\sum_{s_0} ( M^{[0]s_0})^\dagger M^{[0]s_0}= \mathbb{I}_{\chi_1} \Longleftrightarrow \sum_{s_0,k}\overline{(M^{[0]s_0})_{ki}}(M^{[0]s_0})_{kj} =
\sum_{s_0}\overline{U_{s_0i}}U_{s_0j} = \delta_{ij}, \quad 1\leq i,j\leq \chi_1
\end{eqnarray}
due to the unitary matrix $U$ where $\mathbb{I}_{\chi_1}$ is the $\chi_1\times \chi_1$ identity matrix. Here the
summation over $k=1$ can be ignored. Since $\{M^{[i]}\}_{i=0}^{N-2}$ satisfy the left canonical condition, one has
\begin{eqnarray}\label{23.9.20.10}
\langle \psi |\psi \rangle
    &=&  \sum_s |\psi_{s_{0},\cdots,s_{N-1}}|^2   \nonumber \\
 &=& \sum_s (M^{[0]s_0})_{1,\beta_1} \ (M^{[1]s_1})_{\beta_1\beta_2}   \cdots (M^{[N-1]s_{N-1}})_{\beta_{N-1},1}
 \nonumber \\
 && \hspace{0.6cm} \overline{(M^{[0]s_0})_{1,\gamma_1}} \ \overline{(M^{[1]s_1})_{\gamma_1\gamma_2}}  \cdots
 \overline{(M^{[N-1]s_{N-1}})_{\gamma_{N-1},1}} \nonumber \\
&=&\sum_{s_{N-1},\alpha_{N-1}}\overline{(M^{[N-1]s_{N-1}})_{\alpha_{N-1},1}}(M^{[N-1]s_{N-1}})_{\alpha_{N-1},1} \nonumber \\
&=&
\sum_{s_{N-1},\alpha_{N-1}}\overline{ S_{\alpha_{N-1}} V^\dagger_{\alpha_{N-1},s_{N-1}} }  S_{\alpha_{N-1}}V^\dagger_{\alpha_{N-1},s_{N-1}} \\
 &=&
\sum_{s_{N-1},\alpha_{N-1}}  S_{\alpha_{N-1}}^2 V_{s_{N-1},\alpha_{N-1}}  \overline{ V_{s_{N-1},\alpha_{N-1}}}
= \sum_{\alpha_{N-1}=1}^{\chi_{N-1}} S_{\alpha_{N-1}}^2
 \end{eqnarray}
If $\sum_{s_0,\cdots,s_{N-1}}|\psi_{s_{0},\cdots,s_{N-1}}|^2=1$, the left canonical condition for
$M^{[N-1]}$ is also satisfied.

{}


\clearpage

\begin{thebibliography}{}
\bibitem{Feynman_262} R. P. Feynman, Atomic Theory of the Two-Fluid Model of Liquid Helium, Phys. Rev. 94, 262 (1954).

\bibitem{Feynman_102} R. P. Feynman and M. Cohen, Energy Spectrum of the Excitations in Liquid Helium, Phys. Rev. 102, 1189 (1956)

\bibitem{Tsui_1559} D. C. Tsui, H. L. Stormer and A. C. Gossard, Two-dimensional magneto-transport in the extreme quantum limit, Phys. Rev. Lett. 48, 1559 (1982).

\bibitem{Laughlin_1395} R. B. Laughlin, Anomalous quantum Hall effect: An incompressible quantum
fluid with fractionally charged excitations, Phys. Rev. Lett. 50, 1395 (1983).

\bibitem{Haldane_464} F. D. M. Haldane, Continuum dynamics of the 1-D Heisenberg antiferromagnet: Identification with the O(3) nonlinear sigma model, Phys. Lett. A 93, 464 (1983).

\bibitem{Haldane_1153} F. D. M. Haldane, Nonlinear field theory of large-spin Heisenberg antiferromagnets: Semiclassically
 quantized solitons of the one-dimensional easy-axis Neel state, Phys. Rev. Lett.
50, 1153 (1983).

\bibitem{Anderson_153} P. W. Anderson, Resonating valence bonds: A new kind of insulator?, Mater. Res. Bull. 8,
153 (1973).

\bibitem{Bednorz_189} J. G. Bednorz and K. A. Muller, Possible high Tc
superconductivity in the Ba-La-Cu-O system, Z. Physik B - Condensed
Matter 64, 189 (1986).



\bibitem{Daming Li_2016} Fermion bag approach for the massive Thirring model at finite density, Phys. Rev. D 94,
114501 (2016)

\bibitem{White_2863} S. R. White, Density matrix formulation for quantum renormalization groups, Phys. Rev.
Lett. 69, 2863 (1992).

\bibitem{Fannes_443} M. Fannes, B. Nachtergaele and R. F. Werner, Finitely correlated states on quantum spin
chains, Commun. Math. Phys. 144, 443 (1992).

\bibitem{Hastings_2007} M. B. Hastings, An area law for one-dimensional quantum systems, J. Stat. Mech. P08024
(2007).

\bibitem{Schollwck_326} U. Schollw\"ock, The density-matrix renormalization group in the age of matrix product
states, Ann. Phys. 326, 96 (2011).


\bibitem{Liang9214} S. Liang and H. Pang, Approximate diagonalization using the density matrix
renormalization-group method: A two-dimensional-systems perspective, Phys. Rev. B 49,
9214 (1994).

\bibitem{McCulloch852}I. P. McCulloch and M. Gul\'{a}csi, The non-Abelian density matrix renormalization group
algorithm, Europhys. Lett. 57, 852 (2002).

\bibitem{Singh050301} S. Singh, R. N. C. Pfeifer and G. Vidal, Tensor network decompositions in the presence of a
global symmetry, Phys. Rev. A 82, 050301 (2010).

\bibitem{Singh115125} S. Singh, R. N. C. Pfeifer and G. Vidal, Tensor network states and algorithms
in the presence of a global U(1) symmetry, Phys. Rev. B 83, 115125 (2011).

\bibitem{Singh195114} S. Singh and G. Vidal, Tensor network states and algorithms in the presence of a global
SU(2) symmetry, Phys. Rev. B 86, 195114 (2012).

\bibitem{Weichselbaum327} A. Weichselbaum, Non-abelian symmetries in tensor networks: A quantum symmetry space
approach, Ann. Phys. 327, 2972 (2012).


\bibitem{White180403} S. R. White, Density matrix renormalization group algorithms with a single center site,
Phys. Rev. B 72, 180403 (2005).

\bibitem{Hubig155115} C. Hubig, I. P. McCulloch, U. Schollw\"ock and F. A. Wolf, Strictly single-site
DMRG algorithm with subspace expansion, Phys. Rev. B 91, 155115 (2015).

\bibitem{Motruk155139} J. Motruk, M. P. Zaletel, R. S. K. Mong and F. Pollmann, Density matrix renormalization
group on a cylinder in mixed real and momentum space, Phys. Rev. B 93, 155139 (2016).

\bibitem{Ehlers125125} G. Ehlers, S. R. White and R. M. Noack, Hybrid-space density matrix renormalization group
study of the doped two-dimensional Hubbard model, Phys. Rev. B 95, 125125 (2017).

\bibitem{Stoudenmire155137} E. M. Stoudenmire and S. R. White, Real-space parallel density matrix renormalization
group, Phys. Rev. B 87, 155137 (2013).


\bibitem{White076401} S. R. White and A. E. Feiguin, Real-time evolution using the density matrix renormalization
group, Phys. Rev. Lett. 93, 076401 (2004).

\bibitem{Zaletel165112} M. P. Zaletel, R. S. K. Mong, C. Karrasch, J. E. Moore and F. Pollmann, Time-evolving
a matrix product state with long-ranged interactions, Phys. Rev. B 91, 165112 (2015).

\bibitem{Haegeman1103} J. Haegeman, J. Ignacio Cirac, T. J. Osborne, I. Pi\v{z}orn, H. Verschelde and F. Verstraete,
Time-dependent variational principle for quantum lattices, Phys. Rev. Lett. 107, 070601
(2011).

\bibitem{Haegeman165116} J. Haegeman, C. Lubich, I. Oseledets, B. Vandereycken and F. Verstraete, Unifying time
evolution and optimization with matrix product states, Phys. Rev. B 94, 165116 (2016).

\bibitem{Pollmann041116} F. Pollmann, V. Khemani, J. Ignacio Cirac and S. L. Sondhi, Efficient variational diagonalization of fully many-body localized Hamiltonians, Phys. Rev. B 94, 041116 (2016).

\bibitem{Khemani247204} V. Khemani, F. Pollmann and S. L. Sondhi, Obtaining highly excited eigenstates of many-body localized Hamiltonians by the density matrix renormalization group approach, Phys.
Rev. Lett. 116, 247204 (2016).










\bibitem{Yu017201} X. Yu, D. Pekker and B. K. Clark, Finding matrix product state representations of highly
excited eigenstates of many-body localized Hamiltonians, Phys. Rev. Lett. 118, 017201
(2017).

\bibitem{White076401} S. R. White and A. E. Feiguin, Real-time evolution using the density matrix renormalization
group, Phys. Rev. Lett. 93, 076401 (2004).

\bibitem{Zaletel165112} M. P. Zaletel, R. S. K. Mong, C. Karrasch, J. E. Moore and F. Pollmann, Time-evolving
a matrix product state with long-ranged interactions, Phys. Rev. B 91, 165112 (2015).

\bibitem{Haegeman1103} J. Haegeman, J. Ignacio Cirac, T. J. Osborne, I. Pi\v{z}orn, H. Verschelde and F. Verstraete,
Time-dependent variational principle for quantum lattices, Phys. Rev. Lett. 107, 070601
(2011).

\bibitem{Haegeman165116} J. Haegeman, C. Lubich, I. Oseledets, B. Vandereycken and F. Verstraete, Unifying time
evolution and optimization with matrix product states, Phys. Rev. B 94, 165116 (2016).

\bibitem{Pollmann041116} F. Pollmann, V. Khemani, J. Ignacio Cirac and S. L. Sondhi, Efficient variational diagonalization of fully many-body localized Hamiltonians, Phys. Rev. B 94, 041116 (2016).

\bibitem{Khemani247204} V. Khemani, F. Pollmann and S. L. Sondhi, Obtaining highly excited eigenstates of many-body localized Hamiltonians by the density matrix renormalization group approach, Phys.
Rev. Lett. 116, 247204 (2016).

\bibitem{Yu017201} X. Yu, D. Pekker and B. K. Clark, Finding matrix product state representations of highly
excited eigenstates of many-body localized Hamiltonians, Phys. Rev. Lett. 118, 017201
(2017).

\bibitem{Verstraete104} F. Verstraete and J. I. Cirac, Continuous Matrix Product States for Quantum Fields, Phys. Rev. Lett. 104, 190405 (2010).

\bibitem{Haegeman_085118} J. Haegeman, J. Ignacio Cirac, T. J. Osborne and F. Verstraete, Calculus of continuous matrix product states, Phys. Rev. B 88, 085118 (2013)


\bibitem{Rincon_115107} J. Rinc\'{o}n, M. Ganahl, and G. Vidal, Lieb-Liniger model with exponentially decaying interactions: A continuous matrix product state study, Phys. Rev. B 92, 115107 (2015).

\bibitem{Ganahl_220402} M. Ganahl, J. Rincon and Guifre Vidal, Continuous Matrix Product States for Quantum Fields: An Energy Minimization Algorithm, Phys. Rev. Lett. 118, (2017) 220402

\bibitem{Chung_012004} S. S. Chung, S. Bauman, Kuei Sun, and C. J. Bolech, On the new Continuous Matrix Product Ansatz,
J. Phys.: Conf. Ser. 702, 012004 (2016)

\bibitem{Draxler_045145} D. Draxler, J. Haegeman, F. Verstraete, and M. Rizzi, Continuous matrix product states with periodic boundary conditions and an application to atomtronics,
Phys. Rev. B 95, 045145 (2017)

\bibitem{Tilloy_096007} A. Tilloy, Relativistic continuous matrix product states for quantum fields without cutoff,
Phys. Rev. D 104 (2021) 096007.

\bibitem{Carleo_045002} G. Carleo, I. Cirac, K. Cranmer, L. Daudet, M. Schuld, N. Tishby, L. Vogt-Maranto and
L. Zdeborov$\acute{\text{a}}$, Machine learning and the physical sciences, Rev. Mod. Phys. 91, 045002
(2019)

\bibitem{Carleo_602} G. Carleo and M. Troyer, Solving the quantum many-body problem with artificial neural
networks, Science 355, 602 (2017)

\bibitem{Nomura96} Y. Nomura, A. S. Darmawan, Y. Yamaji, and M. Imada, Neural-Network Quantum States, String-Bond States, and Chiral
Topological States, Phys. Rev. B 96, 205152 (2017).
\bibitem{Gao662} X. Gao and L.-M. Duan, Efficient representation of quantum many-body states with deep neural networks, Nat.
Commun. 8, 662 (2017)

\bibitem{Cai97} Z. Cai and J. Liu, Approximating quantum many-body wave functions using artificial neural networks, Phys. Rev. B 97, 035116 (2018).

\bibitem{Carleo5322} G. Carleo, Y. Nomura, and M. Imada, Constructing exact representations of quantum many-body systems with deep neural networks, Nat. Commun. 9, 5322 (2018).

\bibitem{Freitas16} N. Freitas, G. Morigi, and V. Dunjko, Neural network operations and Susuki-Trotter evolution of neural network states, Int. J. Quantum. Inform. 16, 1840008 (2018).

\bibitem{Glasser011006} I. Glasser, N. Pancotti, M. August, I. D. Rodriguez, and J. I.
Cirac, Neural-Network Quantum States, String-Bond States, and Chiral Topological States, Phys. Rev. X 8, 011006 (2018).

\bibitem{Torlai120} G. Torlai and R. G. Melko, Latent Space Purification via Neural Density Operators
, Phys. Rev. Lett. 120, 240503 (2018)

\bibitem{Zen101} R. Zen, L. My, R. Tan, F. H$\acute{e}$bert, M. Gattobigio, C. Miniatura,
 D. Poletti, and S. Bressan, Transfer learning for scalability of neural-network quantum states
, Phys. Rev. E 101, 053301 (2020).

\bibitem{Nomura96} Y. Nomura, A. S. Darmawan, Y. Yamaji, and M. Imada, Restricted Boltzmann machine learning for solving strongly correlated quantum systems, Phys. Rev. B 96, 205152 (2017).

\bibitem{Reh127} M. Reh, Markus Schmitt, and Martin G\"{a}rttner, Time-Dependent Variational Principle for Open Quantum Systems with Artificial Neural Networks, Phys. Rev. Lett. 127, 230501 (2021)


\bibitem{Vieijra124} T. Vieijra, C. Casert, J. Nys, W. D. Neve, J. Haegeman, J. Ryckebusch, and F. Verstraete, Restricted Boltzmann Machines for Quantum States with Non-Abelian or Anyonic Symmetries, Phys. Rev. Lett. 124, 097201 (2020)

\bibitem{Golubeva105} A. Golubeva and R. G. Melko, Pruning a restricted Boltzmann machine for quantum state reconstruction, Phys. Rev. B 105, 125124 (2022).

\bibitem{Pilati101} S. Pilati and P. Pieri, Simulating disordered quantum Ising chains via dense and sparse restricted Boltzmann machines, Phys. Rev. E 101, 063308 (2020).

\bibitem{Lu99} S. Lu, X. Gao, and L.-M. Duan, Efficient representation of topologically ordered states with restricted Boltzmann machines, Phys. Rev. B 99, 155136
(2019).

\bibitem{Nomura127} Y. Nomura, N. Yoshioka, and F. Nori, Purifying Deep Boltzmann Machines for Thermal Quantum States
, Phys. Rev. Lett. 127,
060601 (2021).

\bibitem{Vicentini122} F. Vicentini, A. Biella, N. Regnault, and C. Ciuti, Variational Neural-Network Ansatz for Steady States in Open Quantum Systems, Phys. Rev. Lett. 122, 250503 (2019).

\bibitem{Yoshioka99} N. Yoshioka and R. Hamazaki, Constructing neural stationary states for open quantum many-body systems, Phys. Rev. B 99, 214306 (2019).

\bibitem{Nagy122} A. Nagy and V. Savona, Variational Quantum Monte Carlo Method with a Neural-Network
Ansatz for Open Quantum Systems, Phys. Rev. Lett. 122, 250501 (2019)

\bibitem{Hartmann122} M. J. Hartmann and G. Carleo, Neural-Network Approach to Dissipative Quantum Many-Body Dynamics, Phys. Rev. Lett. 122, 250502 (2019).

\bibitem{Nomura33} Y. Nomura, Neural-Network Approach to Dissipative Quantum Many-Body Dynamics,J. Phys.: Condens. Matter 33, 174003 (2021).

\bibitem{Park2022} C.-Y. Park and M. J. Kastoryano, Expressive power of complex-valued restricted Boltzmann machines for solving nonstoquastic Hamiltonians, Phys. Rev. B 106, 134437
(2022).

\bibitem{Park2020} C.-Y. Park and M. J. Kastoryano, Geometry of learning neural quantum states, Phys. Rev. Res. 2, 023232
(2020).


 \bibitem{Choo121} K. Choo, G. Carleo, N. Regnault, and T. Neupert, Symmetries and Many-Body Excitations with Neural-Network Quantum States, Phys. Rev. Lett. 121, 167204 (2018).


\bibitem{Cai97} Z. Cai and J. Liu, Approximating quantum many-body wave functions using artificial neural networks
, Phys. Rev. B 97, 035116 (2018).

\bibitem{Luo122} D. Luo and B. K. Clark, Backflow Transformations via Neural Networks for Quantum Many-Body Wave Functions
, Phys. Rev. Lett. 122, 226401 (2019).
 \bibitem{Sharir124}O. Sharir, Y. Levine, N. Wies, G. Carleo, and A. Shashua, Deep Autoregressive Models for the Efficient Variational Simulation of Many-Body Quantum Systems, Phys. Rev. Lett. 124, 020503 (2020).
 \bibitem{Kessler2021}J. Kessler, F. Calcavecchia, and T. D. K\"{u}hne, Artificial Neural Networks as Trial Wave Functions for Quantum Monte Carlo, Adv. Theor.
Simul. 4, 2000269 (2021).


\bibitem{Hibat2020} M. Hibat-Allah, M. Ganahl, L. E. Hayward, R. G. Melko, and
J. Carrasquilla, Recurrent neural network wave functions, Phys. Rev. Res. 2, 023358 (2020).

\bibitem{Roth2020} C. Roth, Iterative Retraining of Quantum Spin Models Using Recurrent Neural Networks, arXiv:2003.06228 (2020).

 \bibitem{Choo100}K. Choo, T. Neupert, and G. Carleo, Two-dimensional frustrated $J_1-J_2$ model studied with neural network quantum statesPhys. Rev. B 100, 125124 (2019).

\bibitem{Schmitt125} M. Schmitt and M. Heyl, Quantum Many-Body Dynamics in Two Dimensions with Artificial Neural Networks
, Phys. Rev. Lett. 125, 100503 (2020).

\bibitem{Guti627} I. L. Guti$\acute{e}$rrez and C. B. Mendl, Quantum 6, 627 (2022).

\bibitem{Yang012039}L. Yang, Z. Leng, G. Yu, A. Patel, W.-J. Hu, and H. Pu, Deep learning-enhanced variational Monte Carlo method for quantum many-body physics, Phys.
Rev. Res. 2, 012039(R) (2020).
\bibitem{Irikura013284} N. Irikura and H. Saito, Neural-network quantum states at finite temperature
, Phys. Rev. Res. 2, 013284 (2020).
 \bibitem{Vieijra104}T. Vieijra and J. Nys, Many-body quantum states with exact conservation of non-Abelian and lattice symmetries through variational Monte Carlo
, Phys. Rev. B 104, 045123 (2021).
\bibitem{Liang98} X. Liang, W.-Y. Liu, P.-Z. Lin, G.-C. Guo, Y.-S. Zhang, and L.
He, Solving frustrated quantum many-particle models with convolutional neural networks
, Phys. Rev. B 98, 104426 (2018).
\bibitem{Liu103} C.-Y. Liu and D.-W. Wang, Random sampling neural network for quantum many-body problems
, Phys. Rev. B 103, 205107 (2021).
 \bibitem{Saito87}H. Saito and M. Kato, Machine Learning Technique to Find Quantum Many-Body Ground States of Bosons on a Lattice, J. Phys. Soc. Jpn. 87, 014001 (2018).
 \bibitem{Roth05085}C. Roth and A. H. MacDonald, Group Convolutional Neural Networks Improve Quantum State Accuracy, arXiv:2104.05085 (2021).
 \bibitem{Fu07370} C. Fu, X. Zhang, H. Zhang, H. Ling, S. Xu, and S. Ji, Lattice Convolutional Networks for Learning Ground States of Quantum Many-Body Systems, arXiv:2206.07370 (2021).

\bibitem{Liang2204} X. Liang, M. Li, Q. Xiao, H. An, L. He, X. Zhao, J. Chen, C.
Yang, F. Wang, H. Qian, L. Shen, D. Jia, Y. Gu, X. Liu, and Z.
Wei, $2^{1296}$ Exponentially Complex Quantum Many-Body Simulation via Scalable Deep Learning Method, arXiv:2204.07816 (2022).

\bibitem{Wang394} Y. Wang, Quantum Computation and Quantum Information, statistical Science 2012, Vol. 27, No. 3, 373-394

\bibitem{Li2017} Y. Li and S. C. Benjamin, Efficient variational
quantum simulator incorporating active error minimization, Physical Review X 7, 021050 (2017).

\bibitem{Yuan191} X. Yuan, S. Endo, Q. Zhao, Y. Li, and S. C. Benjamin, Theory of variational quantum simulation, Quantum 3, 191 (2019).

\bibitem{McArdle2019} S. McArdle, T. Jones, S. Endo, Y. Li, S. C. Benjamin, and X. Yuan, Variational ansatz based quantum simulation of imaginary time evolution,
npj Quantum Information 5, 1-6 (2019)

\bibitem{Endo15} S. Endo, J. Sun, Y. Li, S. C. Benjamin,
and X. Yuan, Variational quantum simulation of general processes, Physical Review Letters 125, 010501
(2020).

\bibitem{Yao2011} Y. X. Yao, N. Gomes, Feng Zhang, T.
Iadecola, C. Z. Wang, K. M. Ho, and P. P. Orth, Adaptive variational quantum dynamics
simulations, arXiv preprint arXiv:2011.00622 (2020).

\bibitem{Zhang2011} Z. J. Zhang, J. Sun, X. Yuan, and M. H. Yung, Low-depth hamiltonian simulation by adaptive product formula, arXiv preprint arXiv:2011.05283
(2020)

\bibitem{Heya1904} K. Heya, K. M. Nakanishi, K. Mitarai, and
K. Fujii, Subspace variational quantum simulator, arXiv preprint arXiv:1904.08566 (2019)

\bibitem{Nakanishi2019} K. M. Nakanishi, K. Mitarai, and K. Fujii, Subspace-search variational quantum eigensolver for excited states, Physical Review Research 1, 033062
(2019).

\bibitem{Cirstoiu2020} C. Cirstoiu, Z. Holmes, J. Iosue, L.
Cincio, P. J. Coles, and A. Sornborger, Variational fast forwarding
for quantum simulation beyond the coherence time, npj Quantum
Information 6, 1-10 (2020).

\bibitem{Gibbs2021} J. Gibbs, K. Gili, Z. Holmes, B. Commeau, A. Arrasmith, L. Cincio, P. J.
Coles, and A. Sornborger, Long-time simulations
with high fidelity on quantum hardware, arXiv preprint
arXiv:2102.04313 (2021).

\bibitem{Khatri140} S. Khatri, R. LaRose, A. Poremba, L. Cincio, A. T.
Sornborger, and P. J. Coles, Quantum-assisted quantum compiling, Quantum 3, 140 (2019).

\bibitem{Commeau2009} B. Commeau, M. Cerezo, Z. Holmes, L. Cincio, P. J. Coles, and A. Sornborger,
Variational hamiltonian diagonalization for dynamical quantum simulation, arXiv preprint arXiv:2009.02559
(2020)



\bibitem{Abrams5165} Daniel S. Abrams and Seth Lloyd, Quantum algorithm
providing exponential speed increase for finding eigenvalues and
eigenvectors, Phys. Rev. Lett. 83, 5162- 5165 (1999).

\bibitem{Aspuru1704} A. Aspuru-Guzik, A. D. Dutoi, P. J. Love,
and Martin Head-Gordon, Simulated quantum computation of molecular
energies, Science 309, 1704-1707 (2005).

\bibitem{Kandala2019} A. Kandala, K. Temme, A. D. C$\acute{o}$rcoles, A. Mezzacapo, J. M. Chow, and J. M.
Gambetta, Error mitigation extends the computational reach of a
noisy quantum processor, Nature 567, 491-495 (2019).

\bibitem{Higgott156} O. Higgott, D. Wang, and S. Brierley,
Variational quantum computation of excited states, Quantum 3, 156 (2019).

\bibitem{McClean2017} J. R. McClean, M. E. Kimchi-Schwartz, J. Carter, and W. A. de Jong, Hybrid
quantum-classical hierarchy for mitigation of decoherence and determination of excited states, Physical Review A 95, 042308 (2017).

\bibitem{Nakanishi2019} K. M. Nakanishi, K. Mitarai, and K. Fujii, Subspace-search variational quantum eigensolver
for excited states, Physical Review Research 1, 033062 (2019).

\bibitem{Parrish2019} R. M. Parrish, E. G. Hohenstein, P. L.
McMahon, and T. J. Martinez, Quantum computation of electronic
transitions using a variational quantum eigensolver, Physical Review
Letters 122, 230401 (2019)

\bibitem{Saez2018} A Garcia-Saez and J Latorre, Addressing hard classical problems with adiabatically assisted variational
quantum eigensolvers, arXiv preprint arXiv:1806.02287
(2018).

\bibitem{Cerezo2020} M. Cerezo, Kunal Sharma, Andrew Arrasmith, and
Patrick J Coles, Variational quantum state eigensolver, arXiv preprint arXiv:2004.01372 (2020)

\bibitem{Wang2019} D. Wang, O. Higgott, and S. Brierley, Accelerated variational quantum eigensolver, Physical
Review Letters 122, 140504 (2019).

\bibitem{Wang010346} G. Wang, D. E. Koh, P. D. Johnson,
and Y. Cao, Minimizing estimation runtime on
noisy quantum computers, Phys. Rev. X Quantum 2,
010346 (2021).

\bibitem{Wang09350} G. Wang, D. E. Koh, P. D. Johnson, Y. Cao, Bayesian inference with engineered likelihood functions for robust amplitude estimation, Preprint at https://arxiv.org/abs/2006. 09350 (2020)

\bibitem{Wecker042303} D. Wecker, M. B. Hastings, and M. Troyer, Progress towards practical quantum variational
algorithms, Physical Review A 92, 042303 (2015).

\bibitem{Wiersema_020319} R. Wiersema, C. Zhou, Y. de Sereville, J. F. Carrasquilla, Y. B. Kim,  H. Yuen,
Exploring Entanglement and Optimization within the Hamiltonian Variational
Ansatz, PRX QUANTUM 1 (2020) 020319.

\bibitem{Wen_2019} Wen Wei Ho and Timothy H. Hsieh. Efficient variational
simulation of non-trivial quantum states. SciPost Phys., 6:29, 2019.

\bibitem{Pierre_01862} Pierre-Luc Dallaire-Demers, Michal Stechly, Jerome F. Gonthier,
Ntwali Toussaint Bashige, Jonathan Romero, and Yudong Cao,
An application benchmark for fermionic quantum simulations, arXiv:2003.01862 [quant-ph]

\bibitem{Banuls_158} M. C. Ba$\tilde {\text{n}}$uls, K. Cichy, J. I. Cirac, and K. Jansen, The mass
spectrum of the Schwinger model with matrix product
states, J. High Energy Phys. 11 (2013) 158

\bibitem{Hamer_9701015} C. J. Hamer, Z. Weihong, and J. Oitmaa, Physical Review D 56, 55 (1997), hep-lat/9701015


\bibitem{Roth_05085} C. Roth and A. H. MacDonald, Group convolutional neural networks improve quantum
state accuracy, arXiv:2104.05085.

\bibitem{Cohen_07576} T. Cohen and M. Welling, Group equivariant convolutional networks, arXiv:1602.07576.

\bibitem{Sharir_124} O. Sharir, Y. Levine, N. Wies, G. Carleo and A. Shashua, Deep autoregressive models for
the efficient variational simulation of many-body quantum systems, Phys. Rev. Lett. 124,
020503 (2020)

\end{thebibliography}
\end{document}